\documentclass{article}
\usepackage{graphicx} % Required for inserting images
\usepackage[utf8]{inputenc}
\usepackage[T1]{fontenc}
\usepackage{newunicodechar}
\usepackage[affil-it]{authblk}
\usepackage{setspace}
\newunicodechar{≈}{\approx}
\newunicodechar{ν}{\nu}
\newunicodechar{∆}{\Delta}
\usepackage{abstract}
\usepackage{amsmath}
\usepackage{amssymb}
\usepackage{placeins}
\usepackage{xcolor}
\usepackage{color}

\newcommand{\mycomment}[1]{}
\def\be{\begin{equation}}
\def\ee{\end{equation}}
\def \bea{\begin{eqnarray}}
\def \eea{\end{eqnarray}}
\def \beal{\begin{align}}
\def \eeal{\end{align}}
\def \nn{\nonumber}
\usepackage{braket}
\usepackage{bm}
\usepackage{subcaption}
\usepackage{hyperref}
\usepackage{float}
\newunicodechar{∼}{\sim}
\newunicodechar{₂}{$_2$}
\usepackage[
  backend=biber,
  style=nature,          % biblatex nature style
  sorting=none,
  autocite=superscript,  % use \autocite{} for superscripts (or see \let\cite below)
  url=false, doi=false, eprint=false, isbn=false
]{biblatex}
\AtEveryBibitem{%
  \clearlist{publisher}%    remove Publisher
  \clearlist{location}%     remove Location/Address
  \clearfield{urldate}%
  \clearfield{month}%       remove month like (Nov.)
  \clearlist{language}%     remove "en." etc.
  \clearfield{langid}%      also suppress language tags
  \clearfield{number}%  
  \clearfield{note}
}

\let\cite\supercite

\usepackage{geometry}
\title{Long-Range Fermi-Polaron Blockade in Monolayer MoSe$_2$}
\author{Jonas M. Peterson$^{1,\dagger*}$, Shibalik Lahiri$^{2,\dagger}$, Monique Tie$^{3,4}$, Shibin Deng$^{1}$, Jierong Wang$^{3,4}$, Minxue Wang$^{5}$, Luke Holtzman$^{6}$, James Hone$^{7}$, Takashi Taniguchi$^{8}$, Kenji Watanabe$^{9}$, Tony F. Heinz$^{3,4,*}$, Valentin Walther$^{1,2,*}$, and Libai Huang$^{1,2,5,*}$}
\date{}

\begin{document}

\maketitle
\begin{center}
\begin{it}
\noindent
$^{1}$Department of Chemistry, Purdue University, West Lafayette, Indiana 47907, USA\\
$^{2}$Department of Physics and Astronomy, Purdue University, West Lafayette, Indiana 47907, USA\\[0.25em]
$^{3}$Department of Applied Physics, Stanford University, Stanford, California 94305, USA\\[0.25em]
$^{4}$SLAC National Accelerator Laboratory, Menlo Park, California 94025, USA\\[0.25em]
$^{5}$Elmore Family School of Electrical and Computer Engineering, Purdue University, West Lafayette, Indiana 47907, USA\\[0.25em]
$^{6}${Department of Applied Physics and Applied Mathematics, Columbia University, New York, NY 10027, USA}\\[0.25em]
$^{7}${Department of Mechanical Engineering, Columbia University, New York, NY 10027, USA}\\[0.25em]
$^{8}${Research Center for Materials Nanoarchitectonics, National Institute for Materials Science, 1-1 Namiki, Tsukuba 305-0044, Japan}\\[0.25em]
$^{9}${Research Center for Electronic and Optical Materials, National Institute for Materials Science, 1-1 Namiki, Tsukuba 305-0044, Japan}\\[0.25em]

\noindent
$^{\dagger}$These authors contributed equally to this work.\\
$^{*}${Correspondence: libai-huang@purdue.edu, vwalther@purdue.edu, tony.heinz@stanford.edu, peter735@purdue.edu}

\end{it}
\end{center}
\newpage
\section*{Abstract}
%\color{orange}
Strong optical nonlinearities at the few-photon level are a central goal for quantum photonics, yet they remain difficult to realize in solid-state systems. In doped two-dimensional semiconductors, coupling between excitons and a degenerate Fermi sea gives rise to exciton--Fermi polarons, many-body quasiparticles whose optical response is governed by fermionic correlations. Here, using femtosecond pump--probe transient absorption microscopy, we directly image the spatially resolved nonlinear optical response of exciton--Fermi polarons in monolayer MoSe$_2$. We observe a pronounced spatial suppression of resonant absorption associated with the attractive Fermi polaron, from which we extract an optical blockade radius that is more than ten times larger than that of the neutral exciton. Microscopic analysis indicates that this extended nonlinearity arises primarily from fermion-mediated interactions between exciton--Fermi polarons. Our results establish exciton--Fermi polarons in two-dimensional semiconductors as electrically tunable, strongly interacting optical quasiparticles, and identify them as a promising platform for ultralow-power nonlinear optical devices.

\newpage
\color{black}
\section*{Introduction}

Strong and tunable optical nonlinearities are a key requirement for quantum technologies ranging from single-photon switches to deterministic few-photon gates.\cite{chang_quantum_2014,peyronel_quantum_2012} In most semiconductor platforms, interactions between ground-state excitons are weak and short-ranged, preventing access to the few-photon nonlinear regime under realistic experimental conditions. One route toward stronger optical nonlinearities is to access highly excited excitonic states. Prominent examples include Rydberg excitons in Cu$_2$O and transition-metal dichalcogenides (TMDCs).\cite{kazimierczuk_giant_2014, Thewes2015_PRL, Chernikov2015_NL, walther_giant_2018} Rydberg excitons, which are conceptually analogous to Rydberg atoms, have been shown to exhibit strong interactions, leading to giant Kerr nonlinearities across the high-$n$ Rydberg series in bulk Cu$_2$O~\cite{Morin2022_PRL}, as well as to display the Rydberg blockade effect in both bulk and polaritonic systems~\cite{orfanakis_rydberg_2022, heckotter_asymmetric_2021}. However, their practical implementation in photonic devices is challenged by the weak oscillator strength of highly excited states in bulk crystals and by decoherence- and disorder-induced linewidth broadening in TMDCs, limiting the number of resolvable Rydberg excitons.
 These factors limit scalability and efficient integration into photonic platforms.

An alternative approach to achieving enhanced nonlinearities is to utilize fermionic correlations induced by coupling excitons to a Fermi sea in doped samples. Excitons coupled to a Fermi sea lead to the emergence of a new lower energy ground state known as the attractive exciton-Fermi polaron, which has been known to substantially reshape optical spectra and enhance nonlinear responses.\cite{Sidler2017_NatPhys,Efimkin2017_PRB,Imamoglu2021_CRPhys,HaugKoch_book, schmidt_fermi_2012, massignan_polarons_2025, huang_quantum_2023}. Monolayer TMDCs constitute a promising platform for exploring these many-body optical nonlinearities. Reduced dielectric screening, large exciton binding energies, and strong oscillator strengths fundamentally enhance interaction effects, while electrostatic gating enables deterministic control over the electron density and Fermi energy~\cite{Sidler2017_NatPhys, tan_interacting_2020, muir_interactions_2022, emmanuele_highly_2020, huang_quantum_2023}.

While these studies have clarified the spectral properties of exciton--Fermi polarons, their nonlinear optical response has so far been characterized primarily in the spectral domain. A central open question concerns the spatial structure of this nonlinearity: over what length scales does optical excitation in a doped two-dimensional semiconductor suppress additional resonant absorption, and how is this spatial scale related to the many-body correlations underlying the exciton--Fermi polaron? Addressing this question is essential for understanding the microscopic origin of polaronic nonlinearities and for assessing their relevance in nanoscale photonic systems.

In this work, we use femtosecond pump–probe transient absorption microscopy~(TAM) to investigate the spatiotemporal nonlinear optical response of exciton--Fermi polarons in monolayer MoSe$_2$. We show that an excitonic excitation induces a reorganization of the electronic environment over length scales more than one order of magnitude larger bare exciton size, leading to a pronounced suppression of additional optical excitations and a strongly enhanced nonlinearity. We further demonstrate that this interaction length scale depends sensitively on temperature, reflecting the role of Fermi-sea coherence and polaron linewidth. To understand the microscopic origin of these effects, we develop a variational theory based on the Chevy ansatz. We characterize the leading order nonlinearity and determine how fermion mediated interactions set the characteristic length scale and how it depends on the Fermi energy, exciton density and temperature. Finally, we present a simplified phenomenological model that captures the essential physics. Together, our findings illustrate exciton-Fermi polarons as a viable platform for realizing quantum photonic technologies in 2D semiconductors.

\section*{Results}
 \paragraph{Exciton–Fermi polarons in gated monolayer MoSe₂} We investigate exciton--Fermi polarons in hBN-encapsulated monolayer MoSe$_2$, where an electrostatic back gate allows continuous tuning of the electron density from charge neutrality to a degenerate electron gas (top, Fig.~1a, optical images of the sample in  Extended Data Fig.~1). This gate control provides a well-defined experimental platform in which excitons can be studied both in the absence of free carriers and in the presence of an extended Fermi sea within the same device. Figure~1a shows a schematic of the device geometry used throughout this work.

Linear optical spectroscopy provides a direct probe of how the excitonic response evolves with carrier density and serves as the starting point for identifying exciton--electron correlated states in this system. At charge neutrality, the optical response of monolayer MoSe$_2$ is dominated by a single neutral exciton resonance. Upon electron doping, coupling between excitons and the electron gas leads to a qualitative restructuring of this resonance. As shown in Fig.~1b, the excitonic feature evolves into two distinct spectral branches, corresponding to the attractive and repulsive exciton--Fermi polaron states (AP and RP, respectively). The corresponding splitting is also observed in photoluminescence spectra as shown in Extended Data Fig.~2. At lower gate voltages, additional charges primarily occupy localized trap states, and the excitonic resonance remains largely unchanged. Once electrons begin to populate extended conduction-band states, at a gate voltage of 2~V corresponding to a Fermi level of approximately 5~meV, the energy splitting between these branches increases linearly with gate voltage, as shown in Extended Data Figs.~2.

To theoretically model the polaron states, we consider a single exciton in the Fermi sea interacting with the electrons with a contact interaction of strength $g$, suitable for tightly bound excitons in TMD monolayers~\cite{PhysRevA.83.021603}. The corresponding Hamiltonian is given by
%%%
\be
\hat{H} = \sum_{\bm k}\epsilon^X_{\bm k} x^\dagger_{\bm k} x_{\bm k}
+ \sum_{\bm k}\epsilon^{E}_{\bm k} c^\dagger_{\bm k} c_{\bm k}
+ {g}\sum_{\bm p, \bm k, \bm q} x^\dagger_{\bm p +\bm q -\bm k} x_{\bm p} c^\dagger_{\bm k} c_{\bm q} .
\ee
%%%
Here, $x^\dagger_{\bm k}$ ($x_{\bm k}$) denotes the bosonic creation (annihilation) operator for an exciton with momentum $\bm k$  and mass $m_{X}$ with kinetic energy given by $\epsilon^X_{\bm k}=k^2/(2m_X)$. Similarly, the fermionic creation~(annhilation) operator~ $c^\dagger_{\bm k}$ ($c_{\bm k}$) creates~(annihilates) electronic excitation of mass $m_E$ and kinetic energy given by  $\epsilon^E_{\bm k}=k^2/(2m_E)$. This model accurately captures the formation of the exciton--electron bound state, namely the exciton--trion, characterized by a binding energy $E_B$ and radius $a_T$ in the low-doping limit. At finite carrier density, the exciton--electron interaction, in principle, generates an infinite hierarchy of particle--hole excitations, which collectively contribute to the formation of polaronic quasiparticle states. For light impurities ($m_X \sim m_E$), variational wavefunctions that truncate the particle--hole hierarchy to low order have been shown to provide an accurate description \cite{parish_polaron-molecule_2011, Sidler2017_NatPhys, parish_highly_2013}. This approach shows excellent agreement with exact numerical methods, including diagrammatic Monte Carlo, as well as with the more elaborate non-Gaussian variational ansatz that incorporate higher-order excitations \cite{vlietinck_diagrammatic_2014, qu_efficient_2022}. To model the experiment, we therefore employ the Chevy ansatz, restricted to a single particle--hole excitation over the Fermi sea ~($|\rm FS\rangle$) and describe the Fermi polaron state at momentum $\bm p$~($a^\dagger_{\bm p}$) as
%%%%%
\begin{equation}
a^{\dagger}_{\bm p}|{\rm FS}\rangle =
\phi_{\bm p} x^\dagger_{\bm{p}} \lvert \mathrm{FS} \rangle
+ \sum_{\bm k, \bm q} \phi_{\bm p{\bm k}\bm{q}}\,
x^\dagger_{\bm p+\bm{q}-\bm{k}}
c^\dagger_{\bm {k}} c_{\bm {q}} \lvert \mathrm{FS} \rangle.
\end{equation}
%%%%%%
The variationally determined energy spectrum shows remarkable agreement with the experimental reflectance spectra (Fig.~1b) after accounting for a gate-voltage--dependent shift and correcting for effects due to band-gap renormalization and screening~\cite{Sidler2017_NatPhys}. We identify the onset of the polaronic regime by the linear increase in the splitting of the attractive and repulsive polaron resonances with gate voltage~\cite{Supplement, huang_quantum_2023}, which marks the filling of electronic conduction band states and the formation of a well-defined Fermi sea. %Below this threshold, doped charges primarily occupy localized defect or trap states and do not form a degenerate electron gas, resulting in a negligible shifts of the excitonic resonance (this entence was already in the earlier paragraph). %Above the threshold, the formation of a finite Fermi energy enables the emergence of exciton--Fermi polarons.

The Fermi polaron formation constitutes a local perturbation of the Fermi sea surrounding the exciton. As schematically depicted and theoretically calculated in Fig.~1c, the polaron formation leads to an accumulation of electron density near the exciton (red shaded area), followed by a depletion region extending over longer distances (blue shaded area). The spatial extent of the short-range accumulation scales with the exciton-trion radius $a_T$ whereas the longer-range depletion is set by the inverse of  Fermi momentum $k_{\rm F}^{-1}$.

%The formation of exciton--Fermi polarons in this regime is well described by standard polaron models in which a mobile exciton is dressed by particle--hole excitations of the Fermi sea. For the purposes of this work, we adopt a minimal theoretical description based on the Chevy ansatz (more details in Methods), which captures the essential physics of exciton dressing and reproduces the experimentally observed linear spectra. Figure~1c shows the calculated attractive-polaron wavefunction within this framework, illustrating how the presence of an excitonic impurity locally restructures the surrounding Fermi sea. This real-space redistribution is schematically illustrated in Fig.~1d and highlights the many-body nature of the polaron. While the microscopic wavefunction is spatially localized, the schematic emphasizes that the collective response of the Fermi sea can extend over larger distances than the polaron core itself.
\begin{figure}
    \centering
    \includegraphics[width=0.9\linewidth]{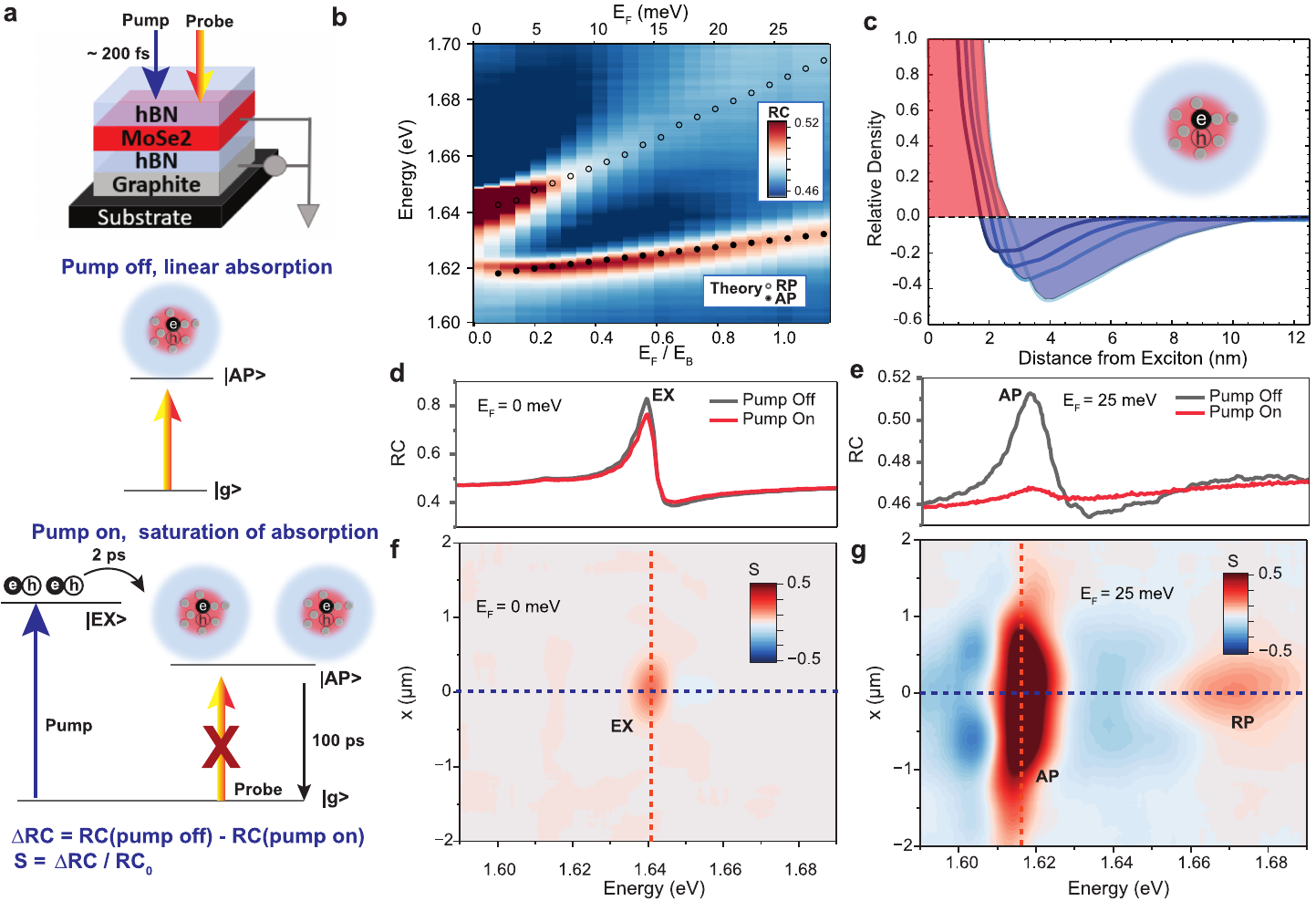}
    \caption{\textbf{Exciton--Fermi polarons and their nonlinear optical response in monolayer MoSe$_2$.} All measurements at 5 K.
\textbf{a,} Schematic of the hBN-encapsulated monolayer MoSe$_2$ device and the experimental measurements. A graphite back gate enables electrostatic tuning of the electron density from the charge-neutral exciton regime to the doped regime where exciton--Fermi polarons form. Linear reflectance spectroscopy identifies the optically active excitonic species as a function of carrier density. The nonlinear and spatially resolved optical response is probed using femtosecond pump–probe transient absorption microscopy (TAM), where a diffraction-limited pump pulse injects excitations slightly above the exciton resonance and a time-delayed, broadband probe measures the pump-induced change in reflectance as a function of energy and position, enabling direct access to the spectral and spatial structure of the nonlinear optical response.
\textbf{b,} Gate-dependent linear reflectance spectra showing the evolution from a single neutral exciton resonance at charge neutrality to two distinct spectral branches corresponding to the attractive and repulsive exciton--Fermi polarons (AP and RP) upon electron doping. Dots represent variationally obtained exciton-Fermi polaron energy using the Chevy ansatz, which are in good agreement with the experimentally observed resonances.
\textbf{c,} The calculated relative change of the local charge density $(n(\mathbf r)-n_{\rm FS})/n_{\rm FS}$ around an exciton fixed at the origin for four representative Fermi energies.The attractive exciton–electron interaction leads to a pronounced short-range accumulation of electronic charge near the exciton, followed by a depletion region at larger distances. The characteristic size of this depletion region increases with decreasing Fermi energy and scales inversely with the Fermi wavevector. 
\textbf{d,e,} Pump-off and pump-on reflectance contrast spectra measured 5~ps after excitation at a pump fluence of approximately $750\,\mu\mathrm{J\,cm^{-2}}$. At charge neutrality (d), excitation of the neutral exciton results in only weak depletion of the absorption. In contrast, excitation of the attractive exciton--Fermi polaron at a Fermi energy of 23~meV (e) leads to strong suppression of the resonant peak amplitude, indicating a strongly enhanced nonlinear optical response.
\textbf{f,g,} Spatially and spectrally resolved TAM response measured 5~ps after excitation under the same conditions as in panels d and e, respectively. For the neutral exciton (f), the nonlinear response remains weak and confined near the pump center. In contrast, excitation of the attractive exciton--Fermi polaron (g) produces a strongly enhanced and spatially extended modification of the optical response. Representative spectral and spatial line cuts are analyzed in Fig.~2.
}
    \label{Fig_1}
\end{figure}

\paragraph{Nonlinear optical response of exciton–Fermi polarons} To probe the nonlinear optical response, we employ femtosecond pump--probe TAM (illustrated in Fig.~1a, more details in Methods and Extended Data Fig.~3). In this technique, a diffraction-limited pump pulse at 710 nm injects excitons slightly above the band gap, while a time-delayed wide-field and broadband probe measures the pump-induced change in reflectance as a function of probe energy and spatial position (Fig.~1d-g). The ultrafast imaging method provides simultaneous access to the evolution of the optical response in energy, space, and time. Following an initial formation period of approximately 2~ps associated with polaron formation, the system enters a quasi-equilibrium regime in which the spatial extent and spectral character of the response remain stable while substantial population loss due to recombination has not yet occurred (Extended Data Fig.~4). Unless otherwise stated, all data shown here are acquired at a pump--probe delay of 5~ps, corresponding to this quasi-equilibrium regime. 

Figures~1d-g show representative TAM measurements acquired at a pump--probe delay of 5~ps, at 5 K. Under identical excitation conditions, the optical response of the neutral exciton and the attractive exciton--Fermi polaron differs strikingly. To quantify the nonlinear optical response, we define a saturation parameter $S = \Delta RC / RC_{0}$, where $\Delta RC$ denotes the pump-induced change in reflectance at the resonance of interest and $RC_{0}$ is the corresponding linear-response amplitude. This parameter measures the reduction of the resonant peak amplitude relative to the linear response. Such a reduction may arise from several spectral modifications, including a rigid shift, broadening, or redistribution of oscillator strength. (Fig.~1d-e). At charge neutrality, the pump-induced reflectance change remains small ($S<0.1$) and is spatially confined to the pump spot (Fig.~1f), even at an exciton density of $4 \times 10^{12}\,\mathrm{cm^{-2}}$, corresponding to an inter-exciton spacing comparable to the exciton Bohr radius. In contrast, in the polaronic regime the response at the attractive-polaron resonance is strongly suppressed (\(S\sim 1\)) at the same pump fluence, and the nonlinear signal extends over a much larger spatial region (Fig.~1g). The observed spatial broadening is consistent with a local saturation mechanism rather than real-space transport, as the transient absorption signal at larger distances from the pump center is nearly indistinguishable from that obtained at lower pump power (Extended Data Fig.~5). This dramatic contrast demonstrates that exciton--Fermi polarons exhibit a much stronger nonlinear optical response than bare excitons, motivating the detailed spatial and spectral analysis presented in the following sections.

 To characterize the nonlinear optical response, we analyze the density-dependent transient reflectance in terms of the saturation parameter $S$. Figure~2a presents the exciton density dependence of $S$ for both the neutral exciton and the attractive exciton--Fermi polaron. We estimate the injected exciton density \(N_{0}\) from the absorption cross section and the incident photon flux (see Methods for details). For the bare exciton, $S$ scales approximately nearly linearly with the injected excitation density over the range studied. At low excitation densities, where saturation effects are negligible, the initial slope of $S$ versus $N_0$ directly measures the suppression area associated with a single excitation, which for neutral excitons is on the order of the exciton size. For attractive polarons, however, this slope exceeds that of neutral excitons by more than two orders of magnitude, indicating much stronger interactions. For neutural excitons, the spectral position of the excitonic resonance slightly blueshifts, and the spatial profile of the response closely follows that of the pump beam, exhibiting only weak broadening with increasing excitation density (Fig.~2b-c). These observations are consistent with a weakly interacting excitonic response. In contrast, the attractive polarons exhibit a qualitatively different behavior (Fig.~2d-e). As \(N_{0}\) increases, the differential reflectance rapidly saturates at excitation densities of $10^{10}\mathrm{cm^{-2}}$, more than two orders of magnitude lower than those required to saturate the bare exciton (Fig.~2a). This saturation is accompanied by a pronounced broadening of the spatial profile, which develops a characteristic flat-top shape at high excitation densities (Fig.~2e). Notably, this strong nonlinear response occurs without a significant density-dependent spectral shift of the AP resonance (Fig.~2d).
 
 \begin{figure}
    \centering
    \includegraphics[width=0.9\linewidth]{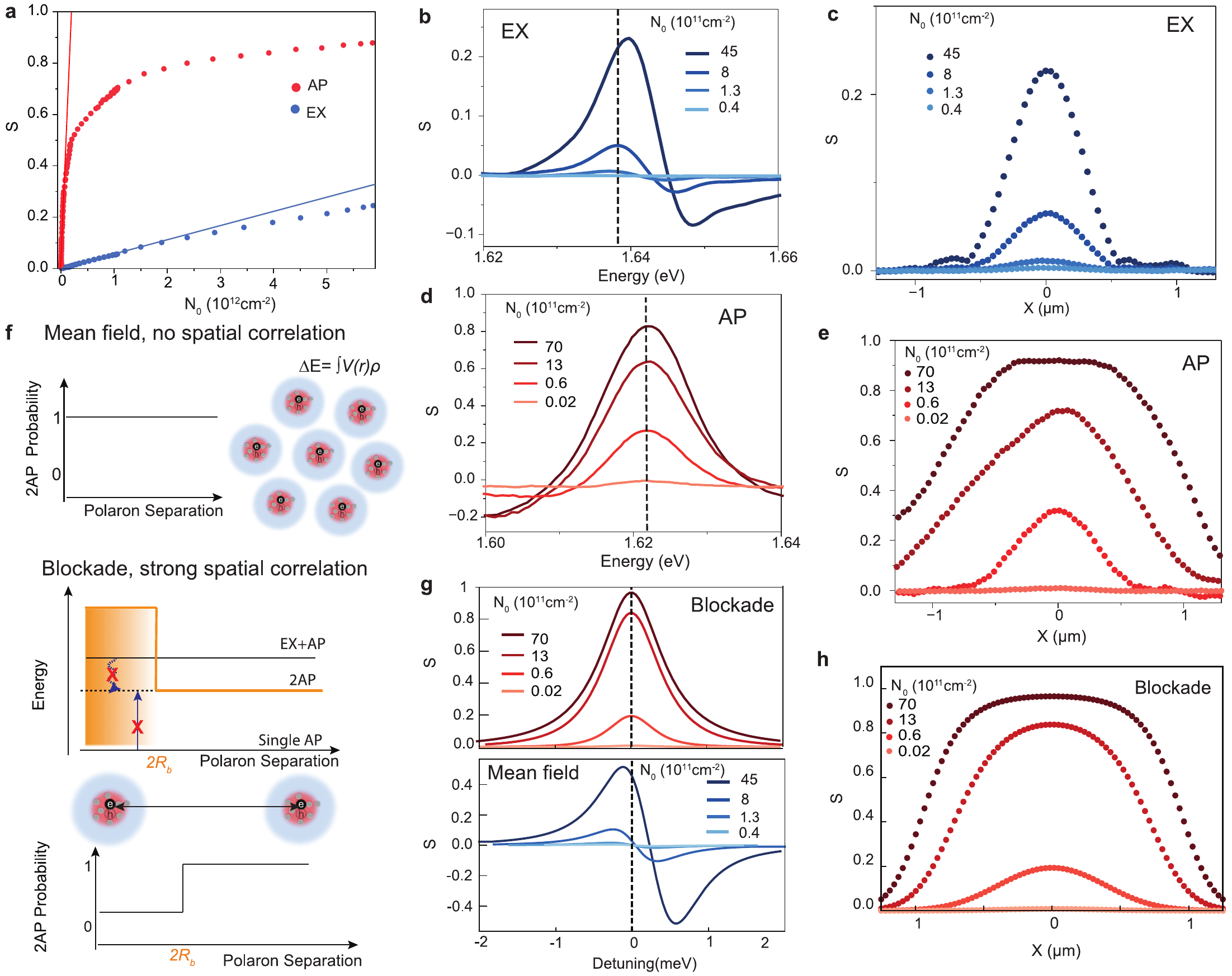}
    \caption{\textbf{Nature of the nonlinear optical response of exciton--Fermi polarons.}
\textbf{a,} Injected-exciton-density ($N_0$) dependence of the saturation parameter. In the low-density regime, the initial slope of $S(N_0)$ provides a direct measure directly measures the suppression area associated with a single excitation. Linear fits to this regime yield a length scale of $1.2 \pm 0.1\,\mathrm{nm}$ for the neutral exciton and $ 14.2 \pm 2.1\,\mathrm{nm}$ for the attractive exciton--Fermi polaron.
\textbf{b,c,} $N_0$-dependent pump-probe transient absorption  response of the neutral exciton. The spectral response (b) exhibits only a small density-dependent blueshift and remains far from full saturation over the studied density range. The corresponding spatial profiles (c) show minimal broadening and closely follow the pump profile, consistent with weak interactions and a mean-field-like response.
\textbf{d,e,} $N_0$-dependent pump-probe transient absorption  response of the attractive exciton--Fermi polaron. The spectral response (d) rapidly saturates with increasing excitation density while exhibiting minimal density-dependent spectral shift. The spatial profiles (e) broaden strongly with increasing density and develop a characteristic flat-top shape at high excitation densities, indicating spatially extended suppression of resonant absorption rather than transport-driven spreading.
\textbf{f,} Schematic comparison of two limiting interaction scenarios. In a mean-field picture (top panel), increasing quasiparticle density primarily leads to energy renormalization and spectral shifts with weak spatial correlations. In the blockade picture (bottom panel), strong inter-polaron correlations define a blockade radius $R_b$, within which an existing pump-generated polaron suppresses both the absorption of probe photons and the relaxation of additional pump-generated excitons into the polaron state.
\textbf{g,} Calculated $N_0$-dependent spectra for the two phenomenological models. A mean-field model~(bottom) produces pronounced spectral shifts inconsistent with experiment, whereas a blockade model~(top) reproduces the saturation of resonant absorption without significant spectral blueshift.
\textbf{h,} Calculated spatial reflectance profile within the blockade model, showing density-dependent broadening and the emergence of a flat-top profile in qualitative agreement with the experimental polaron spatial response. 
}
    \label{fig:placeholder}
\end{figure}

%Further evidence for saturation is provided by the emergence of spectral weight at the repulsive polaron energy once the attractive branch is depleted, indicating that higher-energy polaron states become optically active after the ground state is saturated (Extended Data Fig.~X).

\paragraph{Comparison to interaction models} %The combination of strong saturation, spatial broadening, the emergence of higher-energy polaron features, and the absence of a density-dependent blueshift suggests that the nonlinear optical response of exciton--Fermi polarons cannot be understood within a picture of local or weak interactions. To clarify the origin of this behavior, we compare the experimental observations with theoretical models representing different interaction scenarios (Fig.~2X and details in Methods). A noninteracting picture reproduces the linear excitonic response but fails to account for the observed saturation. Mean-field models with short-range interactions predict a pronounced density-dependent blueshift, in stark contrast to the experimental spectra. By comparison, a blockade-type interaction picture suppresses optical absorption over a finite spatial region without inducing a substantial spectral shift, capturing the key qualitative features observed experimentally.

%\textcolor{blue}{General recommendation on the argumentation line: 1. let's assume there is a polaron-polaron interaction potential $V(r)$. 2. We first assume that the distribution of polarons is uniform, i.e. not much perturbed, even at a finite polaron density. This yield the mean-field result with an effective shift given by $\int V(r) \cdot \rho$ where $\rho$ is the polaron density. 3. This is in stark contrast to the observations. 4. To do better, we account for spatial correlations (such as blockade). This yields the effective potential $V(r)/(V(r) + i \Gamma)$ (or similar). 5. A first reasonable approximation is that the potential has a circular box shape, with strong interactions ($> \Gamma$) at short range and negligible ($<\Gamma$) interactions at large range. -- This way we never have to argue for a contact potential.}
To understand the microscopic origin of the nonlinearity, we compare two phenomenological interaction models between pump- and probe-generated attractive polarons, described by an interaction potential $V(\bm{r})$ (Fig.~2f, top). First, we consider a mean-field  model in which pump-generated polarons are assumed to be uniformly distributed and linearly dependent on the pump intensity. We compute the optical response within the mean-field framework which yields an effective density dependent shift $\int V(\bm r)\rho$, where $\rho$ is the polaron density. Figure 2g displays the differential spectrum computed from this framework showing a density-dependent blue shift which is absent in the attractive polarons (Fig.~2d) and instead more closely resembles the exciton response (Fig.~2b). This discrepancy points to effects beyond the mean-field description for the polarons. 

Next, we consider a blockade model that describes the spatial correlations between the probe-generated and the already present pump-generated attractive polarons via an effective optical potential $V_{\rm eff}(\bm r) = V(\bm r)/(V(\bm r) + i\Gamma/2)$, where $\Gamma$ is the homogeneous linewidth of the AP (see Methods). As a first approximation, we model the interaction potential as a circular box: interactions are strong ($>\Gamma$) at short range and negligible ($<\Gamma$) beyond a blockade radius $R_b$ (Fig.~2f, bottom). Experimentally, in the dilute regime where pump-generated polarons remain well separated, the initial slope of $S$ versus $N_0$ (solid lines in Fig.~2a) directly defines an effective interaction area, $A = \pi R_b^2$. From this analysis, we extract an optical blockade radius of $R_b = 14.2 \pm 2.1\,\mathrm{nm}$ for the polarons. The computed density-dependent spectra using this experimentally determined $R_b$ (Fig.~2g) capture the key features of the attractive polaron response (Fig.~2d), specifically the strong suppression of the resonant peak without a continuous density-dependent spectral shift.

For a weak probe, the total oscillator strength is expected to remain conserved. Within the simplified blockade picture, the total oscillator strength of the probe transition is expected to remain conserved, with some oscillator strength being redistributed to much higher energies. Although the experimental spectra do show evidence of spectral redistribution (Extended Data Figs.~6 and 7), no distinct shifted feature is resolved. This likely reflects the complexity of many-body excited-state absorption at short polaron separations, which is beyond the scope of the simple models used here.

At injected densities \(N_0>1/\pi R_b^2\), the spatial profile develops a flat top near the beam center (Fig.~2e), indicating saturation of the polaron density generated by the pump. Since the pump energy is above the attractive polaron resonance, excitons are first injected and subsequently relax into the polaron state; the observed saturation therefore reflects the inability to relax to generate more polarons (illustrated in Fig.~2f, bottom panel). To model this spatial saturation behavior, we express the local polaron density as:
\begin{equation}
\label{eq_inj_saturation}
\rho(\bm r) = \frac{ N_{\rm 0}(\bm r)}{1 + b\,N_{\rm 0}(\bm r)}.
\end{equation}
Here, \(N_{\rm 0}(\bm r)\) is the above resonance injected exciton density. To obtain $b$, we fit the experimental data in Fig.~2e using Eq.~\eqref{eq_inj_saturation} (Extended Data Fig.~5b). The computed spatial reflectance profile (Fig.~2h) reproduces the observed flat-top and shows qualitative agreement with the experiment. The parameter $b$ captures the saturation and, within the blockade framework, is given by $b = \pi R_b^2$. From fits to the spatial profile at a Fermi level of $23\,\mathrm{meV}$ (Extended Data Fig.~5b), we extract $R_b = 13.1 \pm 2.0\,\mathrm{nm}$, in good agreement with the value $R_b = 14.2 \pm 2.1\,\mathrm{nm}$ obtained from the low-density slope of $S$, confirming that the spatial saturation is governed by similar blockade interactions.

%To understand the nature of the underlying nonlinearity we first model the polaron-polaron interaction via a short ranged contact \textcolor{blue}{(either short-range or contact -- otherwise it's a tautology)} potential \textcolor{blue}{(How the strength of the short-ranged potential chosen?)}. We compute the optical response within a mean-field framework that neglects spatial correlations, yielding the spectrum shown in Fig. 2X with a pronounced density-dependent blue shift. This is in stark contrast with the measured spectra, which display little to no density dependent blue-shift. This discrepancy points to effects beyond a mean-field description. 
% We define this critical separation the blockade radius \textcolor{blue}{(doubling!)} as the separation at which the interaction energy exceeds the linewidth $\Gamma$ of the optical transition. Within this framework, we compute the reflectance for a circular box-shaped interaction potential. Figure 2X shows the obtained spectrum which is in qualitative agreement with the experimental observation of absence of a density-dependent blue shift of the resonance and a reduction in peak reflectance at zero detuning ($\Delta = 0$). This indicates that the nonlinear optical response of exciton--Fermi polarons is governed by finite range interactions going beyond simple mean field description. The spatial extent over which the absorption is suppressed therefore provides insights about the effective interaction length scale, which we analyze quantitatively in the following section.

\paragraph{Interaction-induced blockade radius of exciton--Fermi polarons.} 
%The spatially extended saturation observed in Fig.~2 indicates that resonant optical absorption is suppressed over a finite area surrounding each exciton--Fermi polaron due to strong spatial correlation. To quantify this distance, we extract an effective interaction length scale directly from the density-dependent $S$ in the linear regime. In the dilute excitation regime, where polarons are well separated, the initial slope of $S$ as a function of \(N_{0}\) provides a direct measure of the area over which additional optical excitations are suppressed, $\pi R_{\rm eff}^2$, where $R_{\rm eff}$ is the effective blockade radius. 

We further extend the analysis to various Fermi energies to elucidate the role of the Fermi sea in polaron interactions. Figure~3a compares the low-density slope of $S$ for the attractive polaron at different Fermi energies, from which the corresponding blockade radii $R_b$ are extracted and shown in Fig.~3b. We find that $R_b$ decreases systematically with increasing Fermi energy, ranging from $22.0 \pm 2.9\,\mathrm{nm}$ at $9\,\mathrm{meV}$ to $14.2 \pm 2.1\,\mathrm{nm}$ at $23\,\mathrm{meV}$. This monotonic reduction of the blockade radius with increasing Fermi energy indicates that the nonlinear optical response is governed by interactions mediated by the electron gas rather than by intrinsic properties of the exciton alone.
\begin{figure}
    \centering
    \includegraphics[width=\linewidth]{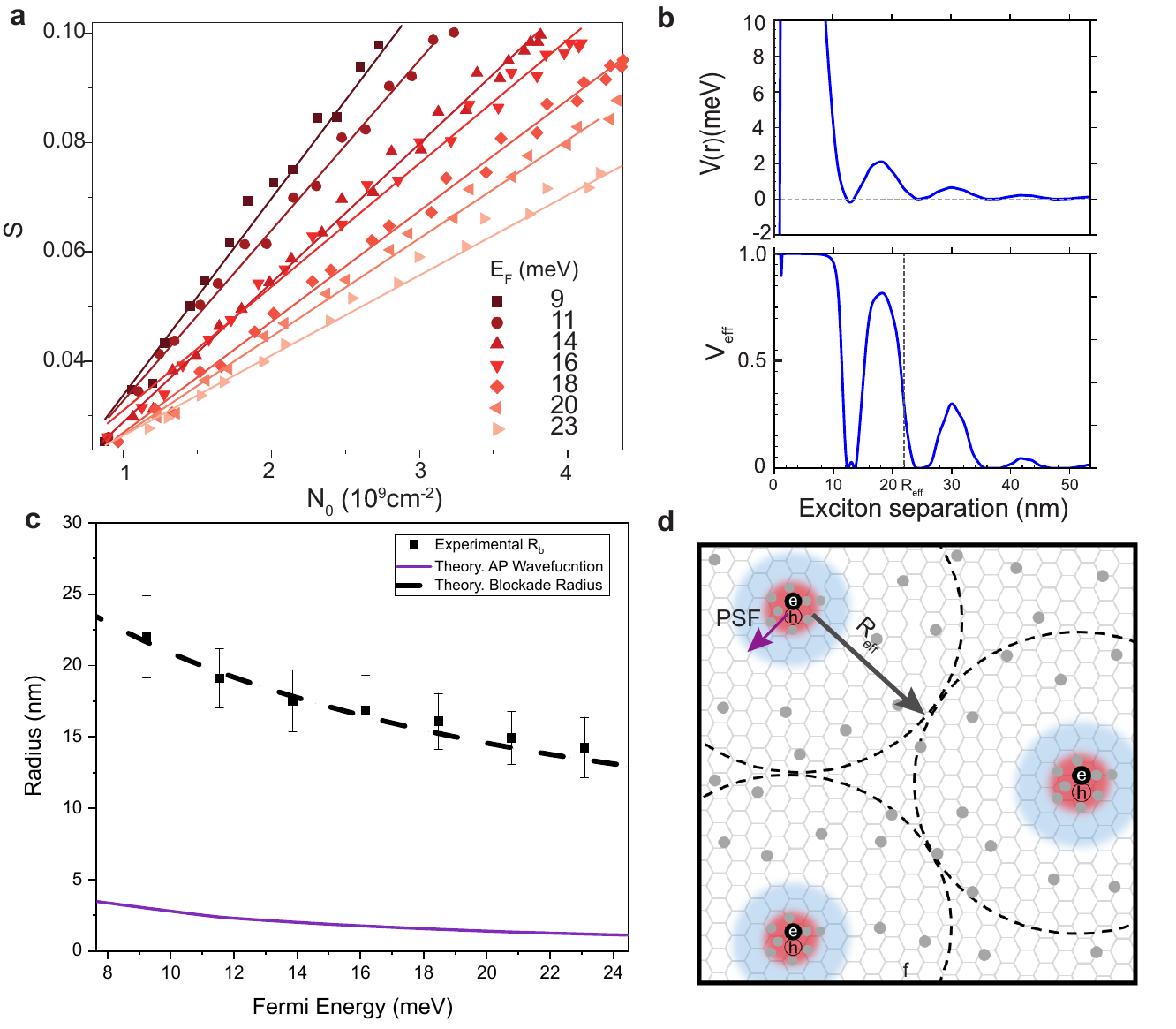}
    \caption{\textbf{Microscopic origin and length scale of the exciton--Fermi polaron blockade.}
\textbf{a,} Optical blockade radius $R_b$ extracted from the low-density slope of the saturation parameter $S$ as a function of injected excitation density, where the initial slope directly measures the suppressed area $\pi R_b^{2}$. The experimentally determined blockade radius decreases monotonically with increasing Fermi energy, ranging from $22.0 \pm 2.9\,\mathrm{nm}$ at $9\,\mathrm{meV}$ to $14.2 \pm 2.1\,\mathrm{nm}$ at $23\,\mathrm{meV}$, indicating a tunable interaction length scale set by the electronic environment. \textbf{b,} Calculated static interaction potential $V(r)$ between excitons mediated by the Fermi sea for a two polaron state in zero momentum mode. Corresponding optically renormalized interaction potential for the given experimental homogeneous linewidth $\Gamma = 1$meV. Within the local density approximation, the integral over this yields the total effective interaction area between optical excitation. 
\textbf{c,} Comparison of the experimentally extracted blockade radius with theoretical estimates as a function of Fermi energy. The purple curve represents the blockade radius due to PSF only, which is significantly smaller than the experimentally obtained interaction range. The black curve represents the obtained interaction-induced blockade radius from the renormalized interaction potential~(\textbf{b}) given a experimental linewidth of ($\Gamma = 1$meV) . This is in close agreement with the experimentally obtained blockade radius~(black squares). \textbf{d,} Schematic highlighting the distinction between the microscopic single-polaron or PSF-associated saturation radius and the much larger interaction-induced optical blockade radius arising direct Coulomb interactions.%~\textcolor{red}{SL: I think fermion mediated is more accurate, since we do not have any coloumb interaction in the system, additionally the exciton-electron interaction which is modeled as contact arises mainly from dipolar interaction}.
}
    \label{fig:placeholder}
\end{figure}

The observed blockade could, in principle, arise from several microscopic interaction mechanisms between polarons, including statistical effects such as Pauli blockade/phase-space filling (PSF) \cite{schmitt-rink_theory_1985, hunsche_exciton_1994} and fermion-mediated interactions \cite{baroni_mediated_2024,giraud_interaction_2012}. We quantify PSF by evaluating the nonlinear change of the optical transition matrix element within the Chevy ansatz {(see Methods)}. From this, we compute the blockade radius associated with a Fermi polaron due to PSF as a function of Fermi energy (Purple solid Fig.~3b). The PSF analysis yields an effective optical blockade radius that is approximately 5--10 times larger than the bare exciton radius. While this already indicates a substantial enhancement of the interaction length scale, it remains smaller than the blockade radius extracted experimentally, underscoring the role of inter-polaron interactions in the measured optical response. 

To estimate the Fermi-sea-mediated interaction between two polarons, we fix the position of one exciton and compute the spatially dependent energy shift of a second static exciton in the zero-momentum polaron mode which is the energetically favored mode that dominates the low-temperature optical response {(see Methods)}. Figure~3c shows the resulting interaction potential, exhibiting short-range attraction together with a long-range repulsive component. The form of potential is 
reminiscent of fermion mediated interactions between heavy impurities systems studied in cold-atom platforms~\cite{ Nishida_Casimir_PRA_2009, Enss_scattering_PRA_2020, baroni_mediated_2024}. To extract an effective optical blockade radius, we evaluate the real part of the optically renormalized interaction potential and equate it to the corresponding interaction area $\pi R_{\rm eff}^2$. Using the measured homogeneous spectral linewidth for the attractive polaron, $\Gamma \approx 1\,\mathrm{meV}$, inferred from our measurements and calibrated using coherence spectroscopy on a closely matched MoSe$_2$ sample~\cite{Moody2015_NatCommun,Martin2016_NanoLett,Jakubczyk2016_PRB},we compute the effective blockade radius shown as dashed gray line in Fig.~3b. The calculated  $R_{\rm eff}$ shows excellent agreement with the experimentally measured  $R_b$  across the full range of Fermi energies studied. This comparison with the effective blockade radius due to PSF (purple line in Fig. 3b) demonstrates that the nonlinear optical response is governed by interaction-induced spatial correlations mediated by the Fermi sea. We emphasize that the present analysis is based on a framework in which only zero-momentum polaron states are retained, neglecting any occupation of finite-momentum modes. This approximation is justified for capturing the leading-order behavior at low excitation densities. Despite its limitation, it provides a microscopic explanation for the pronounced nonlinear optical response observed in exciton–Fermi polarons.

\paragraph{Behavior of repulsive polarons.}
The repulsive exciton--Fermi polaron represents a fundamentally different excitation from the attractive polaron. Unlike the attractive branch, which exhibits pronounced saturation under increasing excitation density, the repulsive polaron does not show clear signs of saturation under the same conditions. Instead, its differential reflectance increases approximately linearly with excitation density (Extended Data Fig.~6 and 7). This contrasting behavior underscores the distinct nature of the repulsive polaron and indicates that its optical response remains unsuppressed even when the attractive polaron absorption has saturated. Such behavior is consistent with the interpretation of the repulsive polaron as an unbound scattering state rather than a bound many-body excitation~\cite{schmidt_fermi_2012}. Finally, the qualitatively different responses of the attractive and repulsive branches argue against pump-induced renormalization effects, such as heating, band-gap renormalization, or dielectric screening, as the origin of the observed saturation, since these mechanisms would be expected to produce broad spectral changes affecting both branches similarly.

%%%%%%%%

\paragraph{Temperature dependence of blockade radius} To further elucidate the origin of the nonlinear optical response of Fermi polarons, we perform temperature-dependent measurements (Fig.~4). With increasing temperature, the most prominent change in the linear optical response is a systematic broadening of the AP resonance, accompanied by a redshift consistent with band-gap renormalization (Fig.~4a). The linewidth increases approximately linearly with temperature, reflecting thermal smearing of the Fermi edge and enhanced scattering processes.

As temperature increases, the nonlinear response of the polarons weakens, with saturation shifting to higher excitation densities. This behavior is quantified in Fig.~4c through the temperature -dependence of the low-density slope of $S$ (Fig~4b). The spatial extent of the nonlinear signal also becomes more confined (Fig.~4c). The extracted $R_{\mathrm{b}}$ decreases monotonically with increasing temperature, indicating a reduced spatial range over which optical absorption is suppressed (Fig.~4d). Over the same temperature range, the neutral exciton exhibits only modest linewidth broadening in the linear spectrum, and the saturation slope remains weakly temperature dependent (Fig.~4e-f). The temperature dependence of $R_{\mathrm{b}}$ follows naturally from the interaction-induced blockade framework: since the effective interaction length scale is limited by the homogeneous polaron linewidth, the systematic linewidth broadening with increasing temperature directly reduces $R_{\mathrm{b}}$, in agreement with the experimental trend (Fig.~4d). A more complete microscopic description of temperature-dependent polaron--polaron interactions would require an explicit finite-temperature treatment, as in Ref.~\cite{Tiene_crossover_PRB2023}.

\begin{figure}
    \centering
    \includegraphics[width=\linewidth]{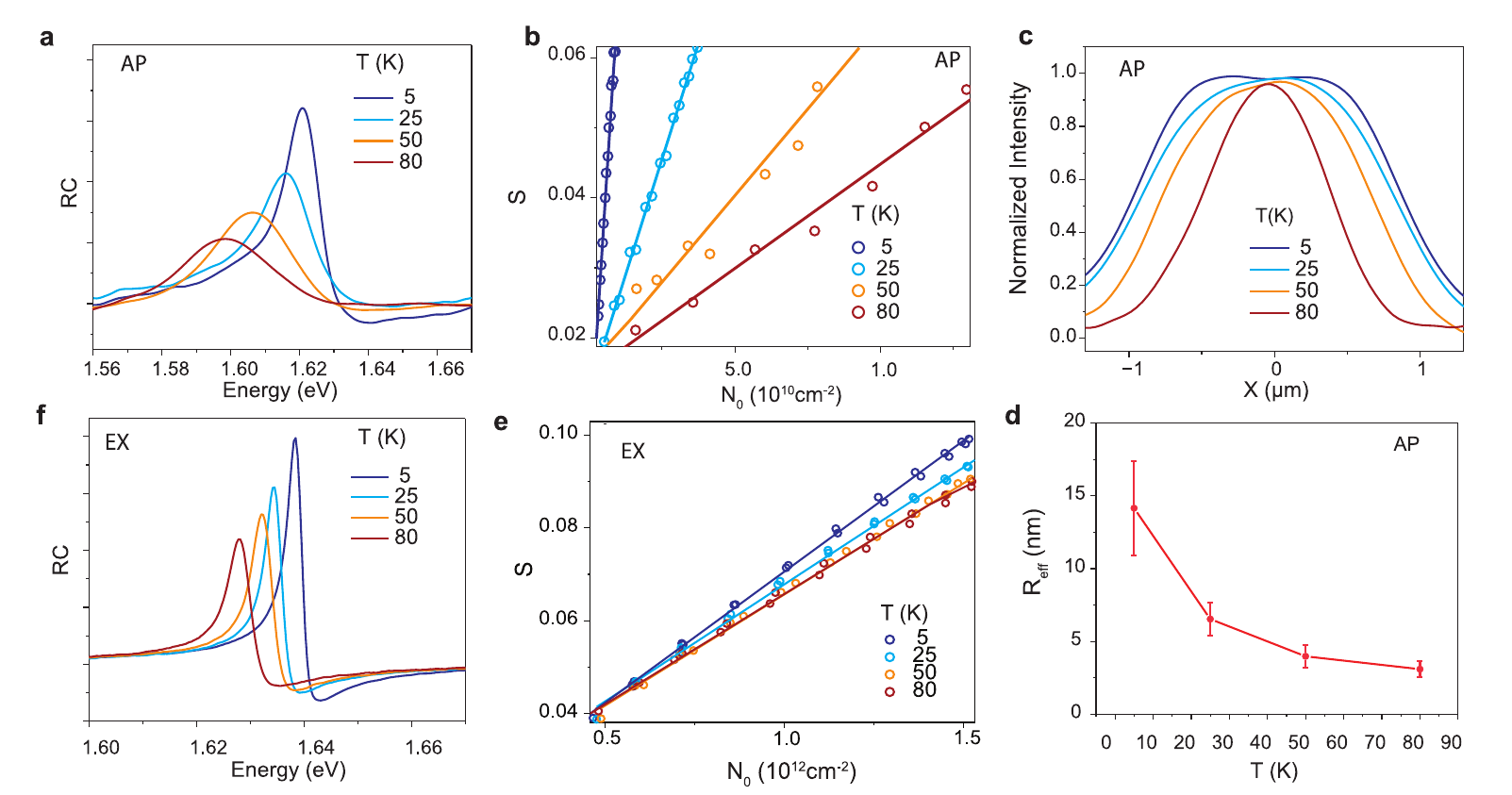}
    \caption{\textbf{Temperature dependence of the blockade radius.}
\textbf{a,} Temperature-dependent linear reflectance spectra of the attractive exciton--Fermi polaron. Increasing temperature leads to a redshift of the resonance, consistent with band-gap renormalization, and a systematic broadening of the polaron spectral linewidth.
\textbf{b,} Change in saturation parameter with density for 4 different sets of temperature for attractive polarons. From the low-density slope the extracted  blockade radius decreases monotonically with increasing temperature, consistent with a reduced interaction range set by the increasing homogeneous linewidth.
\textbf{c,} Spatially resolved transient absorption microscopy response of the attractive exciton--Fermi polaron measured at fixed excitation density for different temperatures. The spatial extent of the nonlinear response decreases with increasing temperature, indicating a reduced saturation area.
\textbf{d,} Optical blockade radius as a function of temperature. The attractive exciton--Fermi polaron exhibits a systematic reduction in blockade radius with increasing temperature.
\textbf{e,} Change in saturation parameter with density for 4 different sets of temperature for excitons. In contrast to the polaron, the slope is only weakly dependent on temperature.
\textbf{f,} Temperature-dependent linear reflectance spectra of the neutral exciton, showing a similar redshift and only modest linewidth broadening over the same temperature range.}
    \label{Fig_4}
\end{figure}

\section*{Discussion}
In this work, we have shown that charge doped monolayer TMDCs are a tunable platform for realizing a strong optical nonlinearity. By using pump-probe TAM imaging and a variational technique for theoretical modeling, we show that exciton-polarons can exhibit blockade radii about 20 times larger than bare excitons. We demonstrate that the optical nonlinearity displays a spatial blockade of excitations analogous to previous studies conducted for Rydberg excitons\cite{orfanakis_rydberg_2022, walther_giant_2018, Kazimierczuk2014_Nature}. However, in contrast to Rydberg-exciton platforms where strong nonlinearities typically rely on accessing high-lying excited states with rapidly diminishing oscillator strength, exciton--Fermi polarons display a sharp increase in nonlinearity without populating higher excited states. Our work is restricted to the 1s excitons-Fermi polaron and one can hypothesize that low-lying Rydberg exciton-Fermi polarons can present even stronger nonlinearity due to long range charge dipole interactions~\cite{liu_exciton-polaron_2021}. 

External electrostatic gating provides a versatile means of tuning the optical blockade radius of exciton--Fermi polarons by controlling the electron density and the resulting many-body correlations. Within this framework, the blockade radius is expected to increase as the homogeneous polaron linewidth is reduced, for example through operation at lower temperatures and in regimes of reduced disorder, and in principle can approach length scales on the order of 100~nm. At very low doping densities, the trion becomes the lowest-energy optical excitation~\cite{qu_efficient_2022}, and in this closely related context unconventional photon blockade based on saturation nonlinearities has been proposed for trion polaritons~\cite{Kyriienko2020_PRL} 

Theoretically, we use the Chevy ansatz to compute the spatially resolved, fermion-mediated static interaction potential between excitons, which provides insights into the origin of the polaronic nonlinearity that underlies the experimentally observed optical blockade. From this interaction potential, we define an effective blockade radius, which shows good agreement with the radius extracted from the measurements. Previous studies in atomic Bose–Fermi mixtures have reported an overall attractive interaction between Fermi polarons~\cite{baroni_mediated_2024}. We note that approximate polaron-polaron interactions we compute are repulsive at large distance, but turn attractive at very short distances. The main effect of this interaction is to blockade further polaron excitation. While this theory yields excellent agreement with our spectroscopic observations, the experiments alone do not allow any conclusions about the sign of the polaron--polaron interaction, which is under debate for Bosonic impurities solid-state \cite{muir_interactions_2022, Tan2023},  and atomic \cite{baroni_mediated_2024} Fermi systems.  However, in atomic platforms the extreme mass imbalance between the Bose and Fermi components leads to a breakdown of the Chevy ansatz due to the importance of higher-order particle–hole excitations, placing those systems in a qualitatively different regime from the solid-state platform studied here. We emphasize that the present theory captures only the leading contributions to the polaron nonlinearity. A full microscopic description, including dynamical and higher-order correlation effects, remains an important direction for future work.

Finally, the interaction-induced blockade radius identified here are comparable to the lateral dimensions of deeply subwavelength nanophotonic and plasmonic structures~\cite{chikkaraddy_single-molecule_2016, Torma__2014}. In such geometries, a single exciton--Fermi polaron could influence the optical response over a substantial fraction of a confined mode, suggesting potential routes toward ultralow-power nonlinear optical devices. While the present experiments probe a classical nonlinear regime, the identification of a mesoscopic saturation length scale establishes a key prerequisite for future studies of photon-number-dependent absorption and modified photon statistics in confined optical environments.

\newpage
\section*{Methods}
\color{black}
\subsection*{Sample Assembly and Characterization}
For this work, high-quality hBN crystals are provided by the National Institute for Materials Science, and low-defect-density MoSe$_2$ crystals are supplied by the Hone Group at Columbia University. All flakes, including graphite, hBN, and MoSe$_2$, are obtained via mechanical exfoliation using Scotch tape. Each layer of the device stack is picked up using the standard PC dry transfer method under ambient conditions~\cite{Zomer2014FastPickup}. The assembled stack is then released onto a clean quartz substrate at 180~$^\circ$C, after which residual PC is removed by sequential immersion in chloroform, acetone, and IPA. The thickness of each layer in the assembled device stack is measured using atomic force microscopy (AFM).

Metal contacts are fabricated via lithography, using either electron-beam lithography (Raith Voyager, 50~keV) with a 200~nm PMMA resist or photolithography (ML3 MicroWriter) with a 1~$\mu$m AZ1512 resist layer. The metal layer (5~nm Ti/50~nm Au) is deposited after lithography using a Kurt J.~Lesker high-vacuum electron-beam evaporator.

\subsection*{Steady-state photoluminescence spectroscopy}

Time-integrated photoluminescence (PL) spectra were measured at $5~\mathrm{K}$ with the sample mounted in a temperature-stabilized closed-cycle cryostat (Montana Instruments). Excitation was provided by an optical parametric amplifier (Light Conversion) producing $\sim 200~\mathrm{fs}$ pulses at a repetition rate of $750~\mathrm{kHz}$ and a wavelength of $710~\mathrm{nm}$. A $100\times$ Zeiss objective (NA $= 0.9$) located inside the cryostat was used both to focus the excitation to a diffraction-limited spot and to collect the epi-emitted PL. The collected emission was directed to an Andor imaging spectrometer and dispersed onto a CCD detector for spectral acquisition. All PL measurements were performed using the same excitation conditions as those employed in the transient absorption microscopy experiments.

\color{black}

\subsection*{Differential reflectance spectroscopy}

Differential reflectance measurements were performed using a collimated tungsten white-light source. The incident optical power was kept low (below a few nanowatts for pulsed measurements) to minimize sample heating and nonlinear effects. The reflected spectrum $R$ from the sample was dispersed by a grating spectrometer and detected using an Andor CCD camera. A reference spectrum $R_0$ was acquired on a nearby region of the device consisting of the same graphite and hBN layers but without the monolayer MoSe$_2$. The reflectance contrast was calculated as $RC = (R - R_0)/R_0$.

\subsection*{Spectral-Resolved Transient Absorption Microscopy (TAM)}
Transient absorption microscopy (TAM) measurements were performed using a home-built pump--probe microscope integrated into a temperature-stabilized closed-cycle cryostat operated at low temperature. The laser source was an ultrafast Yb-based amplifier system (PHAROS, Light Conversion) operating at a repetition rate of 750~kHz with a central wavelength of 1030~nm.

Pump pulses were generated by directing the fundamental output through an optical parametric amplifier to produce excitation pulses at 710~nm with a pulse duration of approximately 200~fs. The pump beam was intensity-modulated using an optical chopper synchronized to the camera acquisition in order to alternate between pump-on and pump-off frames. A high-numerical-aperture microscope objective (100$\times$ NA~=~0.9 Zeiss) mounted inside the cryostat was used to focus the pump beam to a diffraction-limited spot on the sample.

The probe beam was derived from the same 1030~nm fundamental and routed through a motorized mechanical delay stage to control the pump--probe time delay. After the delay stage, the probe beam was focused into a YAG crystal to generate a broadband white-light continuum. The broadband probe was spatially filtered and shaped into a line using a slit, then focused onto the back focal plane of the objective using a wide-field lens, producing a spatially extended probe line on the sample surface.

The probe light reflected from the sample was collected by the same objective, separated from the incident beam using a beam splitter, and directed to a spectrally resolving detection arm. The reflected signal was dispersed by a transmission grating and imaged onto a CMOS camera (Pixelink), such that one camera axis encoded spatial position along the probe line while the orthogonal axis encoded photon energy. Pump-induced changes in the probe reflectance, $\Delta R$, were obtained by subtracting pump-off frames from pump-on frames for each pump--probe delay, yielding simultaneous spatially and spectrally resolved transient reflectance maps.

\subsection*{Determination of exciton density}

We estimate the injected exciton density per pulse from the measured spatial profile of the pump beam and the sample absorptance. The pump spot is characterized by fitting a one-dimensional line cut through the focus to a Gaussian intensity profile, from which the beam radius $w$ (defined by the $1/e^2$ decay of the intensity) is extracted. Assuming a circularly symmetric pump spot, the corresponding two-dimensional fluence profile is
\begin{equation}
F(r) = F_0 \exp\!\left(-\frac{2r^2}{w^2}\right),
\end{equation}
where $F_0$ is the peak fluence at the beam center.

The pulse energy is obtained by integrating the fluence over the focal plane,
\begin{equation}
\label{eq:pulse_energy}
E \;=\; \iint F(r)\,d^2r \;=\; \frac{\pi w^2}{2}\,F_0.
\end{equation}
Independently, the pulse energy is determined from the measured average pump power $P$ and the laser repetition rate $f$ as
\begin{equation}
E \;=\; \frac{P}{f}.
\end{equation}
For an average power of $P = 2.8~\mathrm{nW}$ and a repetition rate of $f = 750~\mathrm{kHz}$, we obtain
\begin{equation}
E \;=\; 3.73\times 10^{-15}\ \mathrm{J}.
\end{equation}

Using the fitted beam radius $w = 0.45~\mu\mathrm{m}$ and Eq.~\eqref{eq:pulse_energy}, the peak fluence is
\begin{equation}
F_0 \;=\; \frac{2E}{\pi w^2}
\;\approx\; 1.17~\mu\mathrm{J\,cm^{-2}}.
\end{equation}
At a pump wavelength of $\lambda = 710~\mathrm{nm}$, the photon energy is
\begin{equation}
E_\gamma \;=\; \frac{hc}{\lambda}
\;=\; 2.80\times 10^{-19}\ \mathrm{J},
\end{equation}
yielding a peak photon areal density
\begin{equation}
N_\gamma(0) \;=\; \frac{F_0}{E_\gamma}
\;\approx\; 4.19\times 10^{12}\ \mathrm{cm^{-2}\,pulse^{-1}}.
\end{equation}
Taking a sample absorptance of $A = 0.0015$ (0.15\%), the injected exciton density at the beam center is
\begin{equation}
n_{\mathrm{ex}}(0) \;=\; A\,N_\gamma(0)
\;\approx\; 6.3\times 10^{9}\ \mathrm{cm^{-2}\,pulse^{-1}}.
\end{equation}

The spatial dependence of the injected exciton density follows directly from the measured Gaussian pump profile,
\begin{equation}
n_{\mathrm{ex}}(x) = n_{\mathrm{ex}}(0)\exp\!\left[-\frac{2(x-x_c)^2}{w^2}\right],
\end{equation}
which yields the exciton density as a function of position along the measured line cut.

\subsection*{Determination of Doping Level and Charge Density}

The carrier density in the monolayer MoSe$_2$ was estimated using a parallel--plate capacitor model based on the graphite back gate. In this device geometry, the graphite gate is separated from the MoSe$_2$ monolayer by a bottom hexagonal boron nitride (hBN) dielectric layer, while the top hBN layer serves only as encapsulation and does not contribute to electrostatic gating.

The gate capacitance per unit area is given by
\begin{equation}
C_g = \frac{\varepsilon_0 \varepsilon_{\mathrm{hBN}}}{t_b},
\end{equation}
where \(\varepsilon_0\) is the vacuum permittivity, \(t_b = 27~\mathrm{nm}\) is the thickness of the bottom hBN layer measured by atomic force microscopy, and the relative dielectric constant of hBN is taken to be \(\varepsilon_{\mathrm{hBN}} = 3.4\)\cite{hBN_dielectric_properties}, consistent with reported values for hBN.

The induced two--dimensional carrier density in the conduction band is then
\begin{equation}
n = \frac{C_g (V_g - V_0)}{e},
\end{equation}
where \(V_g\) is the applied back--gate voltage, \(V_0\) is the gate voltage corresponding to the onset of conduction band filling, and \(e\) is the elementary charge. Below \(V_0\), injected charges predominantly occupy localized states and do not contribute to a degenerate electron gas.

Using these parameters, the carrier density varies with gate voltage at a rate of
\begin{equation}
\frac{dn}{dV_g} \approx 6.9 \times 10^{11}~\mathrm{cm}^{-2}\,\mathrm{V}^{-1}.
\end{equation}

For gate voltages exceeding \(V_0\), a degenerate two--dimensional electron gas forms in the conduction band. The corresponding Fermi energy, measured relative to the conduction band edge, is estimated assuming a parabolic band dispersion as
\begin{equation}
E_F - E_C = \frac{2\pi \hbar^2}{g\,m_e}\,n ,
\end{equation}
where \(n\) is the total two--dimensional electron density, \(m_e\) is the electron effective mass, \(g\) is the total degeneracy factor accounting for spin and valley degrees of freedom, and \(\hbar\) is the reduced Planck constant. In this work, we take \(g = 2\), corresponding to occupation of a single spin--split conduction subband in the two inequivalent valleys, and use an effective electron mass \(m_e = 0.8\,m_0\), where \(m_0\) is the free electron mass.

This procedure enables a direct conversion from gate voltage to carrier density and Fermi energy, which is used to analyze the gate--dependent evolution of excitonic and polaronic resonances.

\subsection*{Chevy ansatz}

To describe an exciton coupled to a Fermi sea in a two-dimensional semiconductor, we employ the Chevy ansatz for the polaron state. The Hamiltonian is given by
\be\label{eq_Ham_exciton_fermion}
\hat{H} = \sum_{\bm p}\epsilon^X_{\bm p} x^\dagger_{\bm p} x_{\bm p} + \sum_{\bm k}\epsilon^{E}_{\bm k} c^\dagger_{\bm k} c_{\bm k} + \frac{g}{L^2}\sum_{\bm p, \bm k, \bm q} x^\dagger_{\bm p +\bm q -\bm k} x_{\bm p} c^\dagger_{\bm k} c_{\bm q} .
\ee
Here, $x^\dagger_{\bm p}$ ($x_{\bm p}$) denotes the bosonic creation (annihilation) operator for an exciton with momentum $\bm k$, and $c^\dagger_{\bm k}$ ($c_{\bm k}$) denotes the fermionic creation (annihilation) operator for electrons in the Fermi sea. The exciton kinetic energy for mass $m_X$ is $\epsilon^X_{\bm p} = p^2/(2m_X)$, and the electron kinetic energy for mass $m_E$ is $\epsilon^E_{\bm k} = k^2/(2m_E)$. We model the exciton-electron interaction via a contact interaction of strength $g$ in a finite box of size $L$. For attractive interactions in two dimensions, there exist a bound state with binding energy $E_B$, which is related to the coupling constant via
\be
\frac{1}{g} = -\frac{1}{L^2}\sum_{\bm k}^{\bm k_\Lambda} \frac{1}{E_B - \epsilon^X_{\bm k} - \epsilon^E_{\bm k}} .
\ee
Here we have introduced an ultraviolet cutoff $\bm k_{\Lambda}$, which does not affect physical results; we take the limit $|\bm k_{\Lambda}|\to \infty$ at the end of the calculation~\cite{Sidler2017_NatPhys, tan_interacting_2020}. In general, the exciton--Fermi-sea interaction generates an infinite hierarchy of particle-hole excitations above the Fermi sea $\ket{\rm FS}$. The Chevy ansatz follows from a variational wave function that truncates the particle-hole excitations to first order~\cite{parish_highly_2013, Sidler2017_NatPhys, parish_polaron-molecule_2011}, given by
\be
\ket{\Psi_{\bm p}} = a_{\bm p}^\dagger\ket{\rm FS}
= \phi_{\bm p}\,x^\dagger_{\bm p}\ket{\rm FS}
+ \sum_{\bm k, \bm q}\phi_{\bm p\bm k\bm q}\,x^\dagger_{\bm p - \bm k + \bm q}c^\dagger_{\bm k}c_{\bm q}\ket{\rm FS}.
\ee
%%%
Here we define $a^\dagger_{\bm p}$ as the quasi-bosonic polaron creation operator. The coefficient $\phi_{\bm p}$ is the amplitude for the bare-exciton component, and $\phi_{\bm p\bm k\bm q}$ is the amplitude for a single particle-hole excitation with particle momentum $\bm k$ and hole momentum $\bm q$. For light impurities, where $m_X \sim m_{E}$, the Chevy ansatz has been shown to accurately describe the polaron state. Minimizing the energy functional $\bra{\Psi_{\bm p}}(E - \hat{H})\ket{\Psi_{\bm p}}$ with respect to $\phi_{\bm p}$ and $\phi_{\bm p\bm k\bm q}$ yields the coupled equations
\bea\label{eq_var_phi_0}
\frac{g}{L^2}\sum_{\bm q}  \phi_{\bm p}  + \frac{g}{L^2}\sum_{\bm k, \bm q}\phi_{\bm p\bm k\bm q}= E\phi_{\bm p}
\\\label{eq_var_phi_kq}
E_X(\bm p, \bm k, \bm q)\phi_{\bm p\bm k \bm q} + \frac{g}{L^2}\phi_{\bm p} +\frac{g}{L^2} \sum_{\bm k'}\phi_{\bm p\bm k'\bm q} - \frac{g}{L^2}\sum_{\bm q'}\phi_{\bm p\bm k \bm q'} = E\phi_{\bm p\bm k\bm q}.
\eea
Here we have defined $E_X(\bm p, \bm k, \bm q) = \epsilon^X_{\bm p +\bm k - \bm q} + \epsilon^E_{\bm k} - \epsilon^E_{\bm q}$. Equations~\eqref{eq_var_phi_0} and \eqref{eq_var_phi_kq} can be solved numerically as a matrix equation, as well as analytically, as done in Refs.~\cite{parish_highly_2013, tan_interacting_2020}. Here we adopt the former approach and compute the eigenspectrum by expanding the wave functions in a B-spline basis. In the following we restrict ourselves to the excitation of the $\bm p = 0$ mode. %Figure~\ref{} shows the full eigenspectrum of the Hamiltonian. The lowest-energy ground state corresponds to the attractive polaron, which connects to the two-body trion bound state at low Fermi energy. The high-lying continuum of states corresponds to the repulsive polaron, connecting to scattering states in the few-body limit. Between the attractive and repulsive polarons lies a band of continuum states known as the trion-hole continuum. The polaron spectral function, depicted in Fig.~\ref{}, exhibits a sharp attractive-polaron resonance and broader features associated with the repulsive-polaron continuum.
%%%
\subsection*{Polaron non-linearity}
%%%%
%%%%
To describe the leading-order nonlinearity, we consider the two-polaron state as a product state, $|P^{(2)}\rangle = ((a_0^\dagger)^2/\sqrt{2})\ket{\rm FS}$, and neglect higher-order states $\propto a^\dagger_{\bm k}a^\dagger_{\bm k'}\ket{\rm FS}$. This perturbative picture is consistent with earlier work~\cite{tan_interacting_2020} and captures leading order nonlinearity  at low densities.

%is based on the assumption that the leading-order polaron nonlinearity arises from Pauli blocking (equivalently, phase-space filling) \textcolor{blue}{(I'm not sure about this last sentence: I think the pertubative approach is accurate if the densities are low and the induced spatial correlations weak. Richard computes an interaction shift from this model.)}.

The fermionic composition of the polaron state makes it susceptible to phase-space filling, which leads to saturation of the optical transition matrix element. This can be seen for a dipole transition operator $\hat T = \Omega(a_0 + a_0^\dagger)$, where the optical transition matrix element deviates from ideal bosonic behavior when transitioning from the single-polaron state ($a_0^\dagger\ket{\rm FS}$) to the two-polaron state $\ket{P^{(2)}}$: 
\be
\frac{1}{\sqrt{2}}\langle{\rm FS}| a_0\,\hat T\, a_0^\dagger a_0^\dagger{\rm FS}\rangle
= \sqrt{2}\,\Omega\left(1 + O_{h} + O_{e} + O_{eh}\right).
\ee
Here we have defined $O_{h}$, $O_{e}$, and $O_{eh}$ as the overlaps evaluate to \cite{tan_interacting_2020}
\bea
O_{h} = -\sum_{\bm k, \bm k' , \bm q}|\phi_{\bm k\bm q}|^2|\phi_{\bm k'\bm q}|^2~,
\\
O_{e} = -\sum_{\bm k, \bm q , \bm q'}|\phi_{\bm k\bm q}|^2|\phi_{\bm k\bm q'}|^2~,
\\
O_{eh} =  \sum_{\bm k, \bm k' , \bm q}|\phi_{\bm k\bm q}|^2|\phi_{\bm k'(\bm q -\bm k + \bm k')}|^2~.
\eea
Here, we have suppressed the zero momentum label and defined $\phi_{\bm k\bm q} \equiv \phi_{0\bm k\bm q}$. Physically, each term corresponds to a distinct mechanism by which the fermionic composition affects the transition into the two-polaron state. The terms $O_{h}$ and $O_{e}$ correspond to the probabilities of overlapping holes and electrons, respectively. Such overlaps are forbidden by the Pauli exclusion principle and therefore suppress the overall transition amplitude. The term $O_{eh}$ corresponds to electron-hole exchange between the two polarons (excitons) and enhances the transition. In practice, the probability of hole overlap is largest due to its restricted phase space (up to the Fermi energy) and provides the dominant contribution to saturation of the optical transition. We therefore neglect $O_e$ and $O_{eh}$ and retain only the saturation effect due to $O_h$ ~\cite{tan_interacting_2020}. We define a saturation density $n_{\rm sat} = 1/(L^2 O_h)$ and a corresponding saturation radius $R_{\rm sat} = \sqrt{L^2O_h/\pi}$. 
%%%
%%%%%
%In addition to saturation, a finite densities of polarons induces an energy shift of polaron resonance due to polaron polaron interactions. These interactions arise from Pauli blocking and direct Coulomb interactions. To calculate the strength of interaction, we compute the two-polaron shift in energy relative to non-interacting polarons. Following Ref.~\cite{tan_interacting_2020}, one can compute the change in energy by adding a second polaron in the system  that is $\langle{\rm FS}| a_0a_0Ha_0^\dagger a_0^\dagger|{\rm FS}\rangle/\langle{\rm FS}| a_0a_0a_0^\dagger a_0^\dagger|{\rm FS}\rangle$ $-2E_0$, where $E_0$ is the single polaron energy. This allows us to compute a density dependent shift and hence an interaction strength given by 
%\bea \nn
%\frac{U}{L^2} = -\frac{1}{\left(1 + O_h\right)}\bigg(\frac{2g}{L^2}\sum |\phi_{\bm k'\bm q}|^2\phi^*_{\bar{\bm k}\bm q}\phi_{\bm k\bm q} + \frac{2g}{L^2}\sum \phi^*_0|\phi_{\bm k\bm q}|^2\phi_{\bm k'\bm q}  
\\
%+2\sum \left(\omega_{\bm q -\bm k} + \epsilon_{\bm k} -\epsilon_{\bm q} \right)|\phi_{\bm k\bm q}|^2|\phi_{\bm k'\bm q}|^2 - 2E_0\bigg) -2E_0~.
%\eea 
%%%%%%
%%%
\subsection*{Spatial behavior}
%%%%
For the preceding analysis, we have considered a delocalized excitation of the polaron state in the $\bm p = 0$ momentum mode, without alluding to any real space representation. To study the spatial characteristics of the polaron state, we evaluate the electron density conditional on the excitons' position. To this end, we rewrite the $\bm p=0$ polaron state in terms of an exciton field operator in real space
\be\label{eq_space}
 a^\dagger_0 = \frac{1}{L}\int d\bm r_X \, X^\dagger(\bm r_X)\left(\phi_0 + \sum_{\bm k, \bm q}\phi_{\bm k\bm q}e^{-i(\bm q -\bm k)\cdot\bm r_X}c^\dagger_{\bm k}c_{\bm q}\right)~.
\ee
Here, we have defined $X^\dagger(\bm r_X)$ as the excitonic field creation operator at position $\bm r_X$. We can similarly define $C^\dagger(\bm r)$~($C(\bm r)$) as the electron field creation (annihilation) operator at position $\bm r$, which can be expressed as $C^\dagger(\bm r) = 1/L \sum_{\bm p} c^\dagger_{\bm p} e^{i\bm p\cdot\bm r}$. Using this form, the electronic density of $a_0^\dagger |\rm FS\rangle$ can be expressed as 
\bea 
\langle C^\dagger(\bm r)C(\bm r) \rangle = \frac{1}{L^2}\int d\bm r_X \left(\rho(\bm r -\bm r_X) + n_{\rm FS}\right)~,
\eea 
where $n_{\rm FS}$ denotes the Fermi-sea density constant for a given Fermi energy and $\rho(\bm r -\bm r_X)$ denotes the local change in electron density, given an exciton at coordinate $\bm r_X$. The function $\rho(\bm r)$ is given by
\be 
\rho(\bm r ) = \frac{1}{L^2}\left[ \sum_{\bm k_1,\bm k_2, \bm q}\phi^*_{\bm k_1\bm q}\phi_{\bm k_2,\bm q} e^{i(\bm k_1 -\bm k_2)\cdot \bm r }- \sum_{\bm k,\bm q_1,\bm q_2}\phi^*_{\bm k\bm q_2}\phi_{\bm k\bm q_1}e^{i(\bm q_1 -\bm q_2)\cdot \bm r } + 2{\rm Re}\sum_{\bm k\bm q}\phi_{\bm k\bm q}\phi_0 e^{i(\bm q -\bm k)\cdot \bm r} \right].
\ee 
Each term in the density expression corresponds to a distinct physical process. The first term describes the density perturbation arising from particle excitations, while the second term describes the perturbation of the Fermi-sea density due to hole excitations. The third term describes Friedel-like oscillations in the Fermi sea induced by the presence of the impurity~(exciton). We extend this formalism to the two-polaron state $\lvert P^{(2)}\rangle$ and evaluate the resulting local fermionic density perturbation and energy shift for excitons fixed at specified spatial coordinates.
To obtain a real-space measure of the interaction mediated by the Fermi sea, we define an effective static interaction kernel by evaluating the interaction-induced energy shift in $\bm p=0$ mode while suppressing exciton dynamics. Concretely, we remove the exciton kinetic term from Eq.~\eqref{eq_Ham_exciton_fermion} and evaluate the change in the many-body energy functional between the one- and two-exciton sectors. This construction yields a nonlocal kernel $V(\bm r)$ whose range and sign encode the Fermi-sea--induced interaction responsible for the observed nonlinearity. Specifically, we compute the expectation value

\bea \nn
\frac{1}{2}\langle{\rm FS}|\left( a^2_0\left(\hat H-  \sum_{\bm p}\omega_{\bm p}x^\dagger_{\bm p}x_{\bm p}\right) {a_0^\dagger}^2\right)|{\rm FS}\rangle -2\langle{\rm FS}|\left( a_0\left(\hat H-  \sum_{\bm p}\omega_{\bm p}x^\dagger_{\bm p}x_{\bm p}\right)  a_0^\dagger\right)|{\rm FS}\rangle
\\
=\int d\bm r_Xd\bm r_X' V(\bm r_X -\bm r_X')~,
\eea 
where $V(\bm r_X -\bm r_X') = V_{\rm att}(\bm r_X -\bm r_X') + V_{\rm rep}(\bm r_X -\bm r_X')$ is the effective interaction potential, which can be decomposed into an attractive part
\begin{align}
V_{\rm att}(\bm r)
&=
\frac{2g}{L^2} |\phi_0|^2 \sum
\phi^*_{\bm k_1 \bm q}\phi_{\bm k_2 \bm q}
\left[
e^{i(\bm k_1-\bm k_2)\cdot \bm r}
+2\,\mathrm{Re}\,e^{i(\bm q-\bm k_1)\cdot \bm r}
\right]
+\frac{2g}{L^2} \sum
\phi^*_{\bm k_1 \bm q'}\phi_{\bm k_2 \bm q'}|\phi_{\bm k\bm q}|^2
e^{-i(\bm k_1-\bm k_2)\cdot \bm r}
\nonumber\\
&\quad
+\frac{4g}{L^2}|\phi_0|^3 \sum
\mathrm{Re}\!\left[\phi_{\bm k\bm q}e^{i(\bm q-\bm k)\cdot \bm r}\right]
+\frac{8g}{L^2}
\mathrm{Re}\!\left[
\phi_0^* \sum
|\phi_{\bm k\bm q}|^2\phi_{\bm k'\bm q'}
\left(
e^{-i(\bm q-\bm k)\cdot \bm r}+e^{-i(\bm q'-\bm k')\cdot \bm r}
\right)
\right]
\nonumber\\
&\quad
+4|\phi_0|^2 \sum
\epsilon_{\bm k}|\phi_{\bm k\bm q}|^2
e^{-i(\bm q-\bm k)\cdot \bm r}
+2\sum
\epsilon_{\bm q}\,
\phi_{\bm k\bm q'}\phi^*_{\bm k'\bm q'}\phi^*_{\bm k'\bm q}\phi_{\bm k\bm q}\,
e^{-i(\bm q'-\bm q)\cdot \bm r} \, ,
\end{align}
%%%
and a repulsive part
%%%
\begin{align}
V_{\rm rep}(\bm r)
&=
\frac{2g}{L^2}
\sum
\phi^*_{\bm k_1 \bm q}\phi_{\bm k_2 \bm q'}\phi^*_{\bm k\bm q'}\phi_{\bm k\bm q}
\,e^{-i(\bm q-\bm q')\cdot \bm r}
+\frac{2g}{L^2}
\Bigg[
\phi_0^*
\sum
\phi^*_{\bm k\bm q'}\phi_{\bm k\bm q}\phi_{\bm k'\bm q'}
\,e^{-i(\bm q-\bm q')\cdot \bm r}
+\mathrm{c.c.}
\Bigg]
\nonumber\\[4pt]
&\quad
+2\sum
\epsilon_{\bm k}\,
\phi_{\bm k\bm q'}\phi^*_{\bm k'\bm q'}\phi^*_{\bm k'\bm q}\phi_{\bm k\bm q}\,
e^{-i(\bm q'-\bm q)\cdot \bm r} +4|\phi_0|^2\sum
\epsilon_{\bm q}|\phi_{\bm k\bm q}|^2
e^{-i(\bm q-\bm k)\cdot \bm r} \, .
\end{align} 
For brevity, we omit the summation indices; it is understood that the sums run over all momenta with $|\bm k| > k_{\rm F}$ and $|\bm q| < k_{\rm F}$. As before, we retain only the contribution from hole overlap in the leading-order nonlinearity. Additionally, we discard any term that vanishes as $k_{\Lambda} \to \infty$. One can verify that the retained terms are cut-off independent and therefore give a finite contribution to the potential~\cite{Sidler2017_NatPhys, tan_interacting_2020}. The origin of the attractive interaction potential is mainly due to the Fermi-sea-mediated interaction between the excitons, whereas the repulsive part mainly stems from Pauli blocking associated with the fermionic components. Additionally, corrections arise from the kinetic energy of the electron and hole excitations within the polaron dressing cloud. 
%%%%%
\subsection*{Phenemological pump-probe blockade model}
%%%%%%
To describe the experimental observables, we adopt a phenomenological model to describe the probe reflectance spectra given a density of pump polarons generated using an off-resonant pump. Since, in the experiment the pump polarons are probed in a dynamical quasi-equilibrium state, we consider a finite distribution of pump polarons~($\rho(\bm r)$) that is probed by a weak intensity laser~($\Omega_{p}$). To effectively describe the system we consider the following Hamiltonian
\bea
H = -\Delta \int d\bm r  A^\dagger(\bm r) A(\bm r) + \Omega_{p}\int d\bm r \left(A^\dagger(\bm r) + A(\bm r)\right) + \int d\bm r d\bm r'V(\bm r -\bm r')A^\dagger(\bm r')P^\dagger(\bm r)P(\bm r)A(\bm r)
\eea 
Here we define $A^\dagger(\bm r)$ as the bosonic operator that creates probe polarons at a detuning $\Delta$ and $P^\dagger(\bm r)$ creates independent pump polarons. 
%The pump and the probe states are distinguishable due to loss of coherence and hence we have $[A^\dagger(\bm r), P(\bm r)] =0$ \textcolor{blue}{(I don't understand this sentence)}. 
We consider an interaction potential $V(\bm r)$ between the pump and probe polarons. Since we operate in the weak-intensity regime, we ignore any interaction between the probe polarons.
The equation of motion for the polarons is given by
\bea 
i\partial_t A(\bm r) &=&\left(-\Delta - i\frac{\Gamma}{2}\right)A(\bm r) +\Omega_p + \int d\bm r' V(\bm r -\bm r')P^\dagger(\bm r')P(\bm r') A(\bm r)
\\\nn
i\partial_t P^\dagger(\bm r')P(\bm r') A(\bm r) &=& \left(-\Delta - i\frac{\Gamma}{2}+ V(\bm r' -\bm r)\right)P^\dagger(\bm r')P(\bm r') A(\bm r) + \Omega_p P^\dagger(\bm r')P(\bm r') 
\\\label{ignored_term}
&+& \int d\bm r'' V(\bm r -\bm r'')P^\dagger(\bm r') P^\dagger(\bm r') A(\bm r') A(\bm r'') A(\bm r)
\eea 
We further apply the local density approximation for the pumped state, that is  we take the pump operators out of the spatial integrals involving the interaction potential $V(\bm r)$. This is justified since the spatial profile of the pump, set by the excitation laser, varies on a much longer length scale than the blockade radius. Quantitatively, the typical spot size in experiments is on the order of $1$ $\mu\text{m}$, whereas the extracted blockade radius is $\sim 20$ $\text{nm}$, i.e., two orders of magnitude smaller. This yields the following expression:
\be \label{eq_A_r_solution} 
\langle A(\bm r) \rangle =\frac{\Omega_p}{-\Delta - i\Gamma/2}\left(1 -\left(\int d\bm r' V_{\rm eff}(\bm r')\right)\rho(\bm r)\right)~.
\ee 
Where we have defined the optically renormalized effective potential as
\be 
V_{\rm eff}(\bm r) = \frac{V(\bm r)}{-\Delta - i\Gamma/2 + V(\bm r)}~.
\ee 
For resonant excitation, i. e., $\Delta =0$, we use the optically renormalized potential to define the blockade radius in the main text. For a finite box-potential of strength $V_0$ and range $R_b$, Eq.~\eqref{eq_A_r_solution} yields
\be\label{eq_A_r_solution_finite_box} 
\langle A(\bm r) \rangle =\frac{\Omega_p}{-\Delta - i\Gamma/2} + \frac{\Omega_{p}V_0\pi R_b^2\rho(\bm r)}{-\Delta + i\Gamma/2 + V_0}~,
\ee 
From $\langle A(\bm r)\rangle$, we compute the reflectance for the situation of blockade~($V_0/\Gamma \to \infty$), which is given by 
\be
R(\bm r, \Delta) = \frac{\Gamma^2}{4\Delta^2 + \Gamma^2}\left(1 -\pi R_b^2\rho(\bm r)\right)
\ee 
To connect it with the experimental observable, we compute the differential reflectance which is $\Delta R/R = (R(\bm r, \Delta) -R(\bm r =0, \Delta))/R(\bm r =0, \Delta =0)$, which yields
\be\label{eq_diff} 
\frac{\Delta R}{R} = \frac{\Gamma^2}{4\Delta^2 + \Gamma^2}\pi R_b^2 \rho(\bm r)
\ee 
Eq.~\eqref{eq_diff} suggests that the probe differential reflectance follows the pump spatial profile. Consequently, the saturation in the pump profile is directly manifested in the reflectance spectra.
\color{black}
\printbibliography
\section*{Acknowledgments} This work is mainly supported by U.S. Department of Energy, Office of Basic Energy Sciences through QuPIDC EFRC award DE-SC0025620 (for the transient and time-resolved optical spectroscopy and microscopy work at Purdue). Spectroscopic studies and analysis at SLAC (T.H. and M.T.) were supported by the US Department of Energy, Office of Science, Basic Energy Sciences, Chemical Sciences, Geosciences, and Biosciences Division through the AMOS program, with additional support from the Gordon and Betty Moore Foundation’s EPiQS Initiative through grant number GBMF9462. Sample preparation at SLAC (J.W.) was by supported by the U.S. Department of Energy, Office of Science, Office of Basic Energy Sciences, Materials Sciences and Engineering Division through the QIS program. K.W. and T.T. acknowledge support from the CREST (JPMJCR24A5), JST and World Premier International Research Center Initiative (WPI), MEXT, Japan.
\section*{Author Contributions}
L.H., V.W., and T.H. designed the experiments and theoretical simulations; J.P., S.D., and M.W. carried out the optical measurements; M.T. and J.W. fabricated and characterized samples; S.L. and V.W. carried out and analyzed the simulations; K.W. and T.T. grew the hBN crystals; J.P., S.L., V.W., and L.H. wrote the manuscript with input from all authors. 
 
\section*{Data availability}
All data supporting the findings of this study are available within the paper and its Extended Data. Source data are provided with this paper.
\section*{Code availability}
Codes that support the findings of this study are available upon request. Codes include scripts for data processing and theoretical modelling.  
\setcounter{figure}{0}
\renewcommand{\thefigure}{\arabic{figure}}
\renewcommand{\figurename}{Extended Data Figure}
\newpage
\begin{figure}
    \centering
    \includegraphics[width=\linewidth]{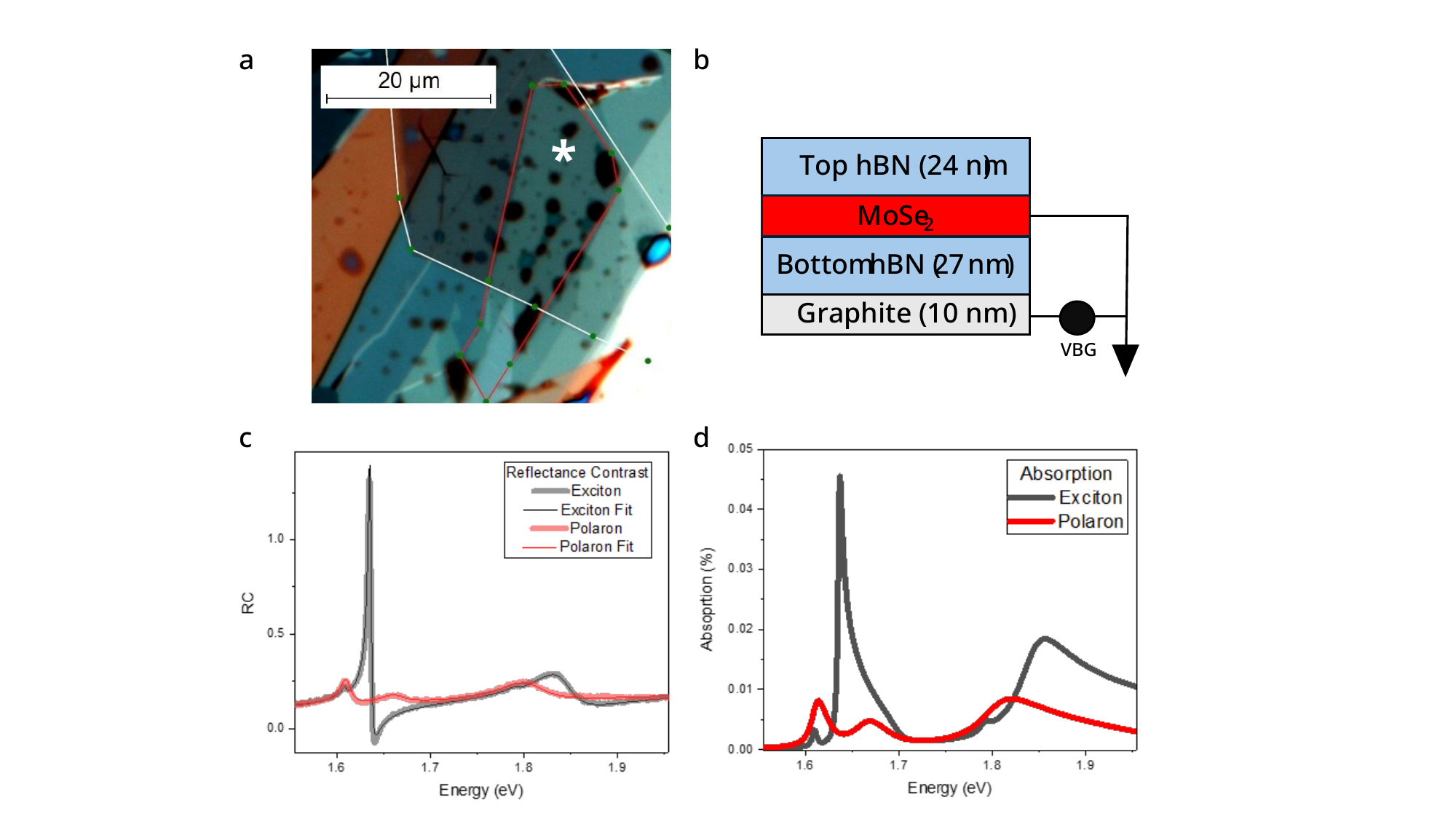}
    \caption{\textbf{Sample characterization and optical response.}
\textbf{a,} Optical image of the hBN-encapsulated monolayer MoSe$_2$ device. The white outline indicates the graphite back gate, and the red outline marks the monolayer MoSe$_2$ region. White asterisks denote the bubble-free region of the sample used for all optical measurements.
\textbf{b,} Illustration of the device layer structure. Individual layer thicknesses were determined using atomic force microscopy and used as inputs for optical modeling of the reflectance and absorption response.
\textbf{c,} Differential reflectance spectra measured at charge neutrality (neutral exciton) and at a Fermi energy of 23~meV (attractive exciton--Fermi polaron). Transparent curves show the measured reflectance contrast, while solid curves are fits using a Fano lineshape.
\textbf{d,} Absorption spectra extracted from the reflectance contrast measurements using a transfer-matrix model. These absorption spectra are used to determine excitonic densities and to quantify the optical response under different doping conditions.
}
    \label{Extendend Data Figure X}
\end{figure}
\begin{figure}
    \includegraphics[width=\linewidth]{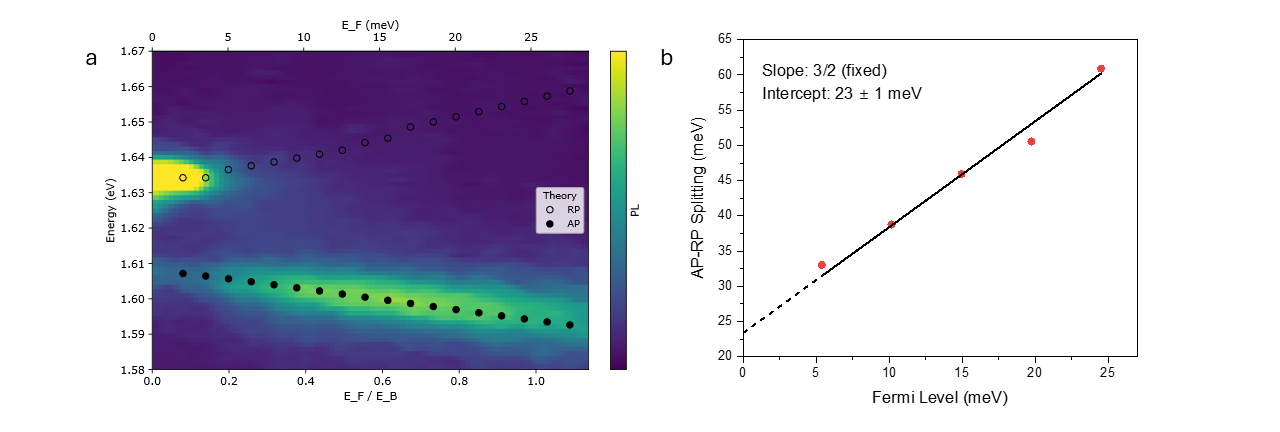}
    \caption{\textbf{Gate-dependent photoluminescence spectra and binding energy extraction.}
\textbf{a,} Gate-dependent photoluminescence (PL) spectra acquired under the same excitation conditions used for the transient absorption microscopy measurements. With increasing electron doping, the PL response evolves from a single neutral exciton emission to two distinct branches corresponding to the attractive (AP) and repulsive (RP) exciton--Fermi polarons, analogous to the behavior observed in linear reflectance. Discrete markers indicate polaron energies calculated using the Chevy ansatz, showing good agreement with the experimentally observed PL peaks.
\textbf{b,} Determination of the polaron binding energy from the separation between the AP and RP emission energies as a function of Fermi energy. The approximately linear dependence of the AP--RP splitting on Fermi energy is extrapolated to zero carrier density, yielding an estimate of the binding energy.
}
    \label{Extendend Data Figure X}
\end{figure}

    \centering
\begin{figure}
    \centering
    \includegraphics[width=\linewidth]{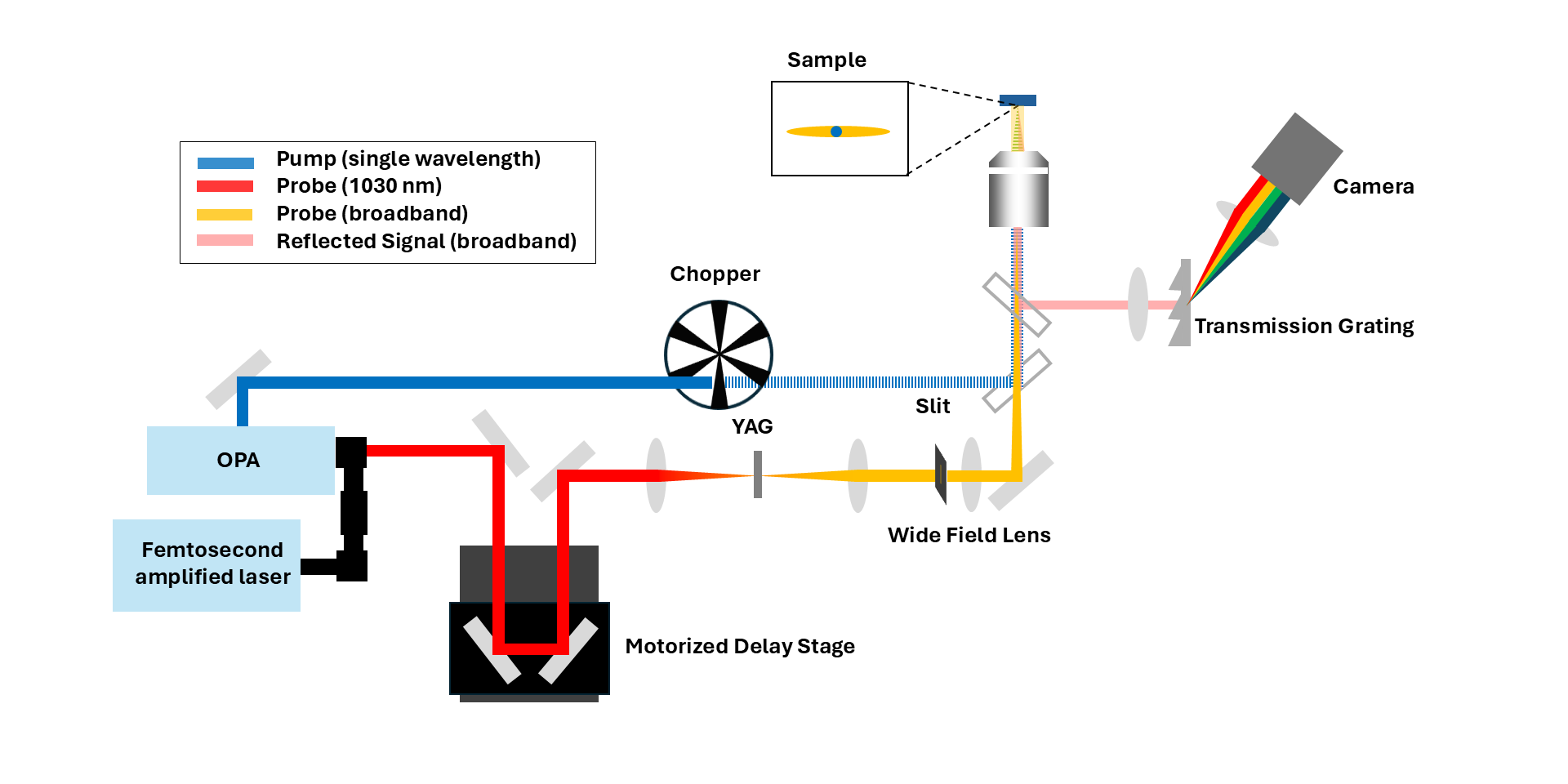}
    \caption{\textbf{Spectrally and spatially resolved transient absorption microscopy setup.}
Schematic of the pump--probe microscope used for spectrally resolved transient absorption microscopy. An ultrafast Yb-based amplified laser system provides femtosecond pulses that are split to generate the pump and probe beams, with the pump wavelength produced using an optical parametric amplifier (OPA). The pump path (blue) is directed through an optical chopper synchronized with the CMOS camera to alternate between pump-on and pump-off measurements and is focused by a high-numerical-aperture objective onto the sample, where it locally generates excitonic quasiparticles.
The probe path (red) is derived from the same amplified laser source and routed through a motorized delay stage, which controls the temporal overlap between pump and probe. The delayed probe is focused into a yttrium aluminum garnet (YAG) crystal to generate a broadband white-light supercontinuum (orange). The broadband probe is shaped by a slit and focused by a wide-field lens onto the back focal plane of the objective, producing a probe that is diffraction-limited along one spatial axis and extended along the orthogonal axis, resulting in a line-shaped probe on the sample.
The reflected probe signal (pink) is collected by the same objective and directed through a transmission grating, which disperses the signal spectrally along one detector axis while preserving spatial information along the orthogonal axis. The pump-induced change in the probe reflectance is recorded on a CMOS camera, enabling simultaneous measurement of temporal, spatial, and spectral components of the transient optical response.
}
    \label{Extendend Data Figure X}
\end{figure}

\begin{figure}
    \centering
    \includegraphics[width=\linewidth]{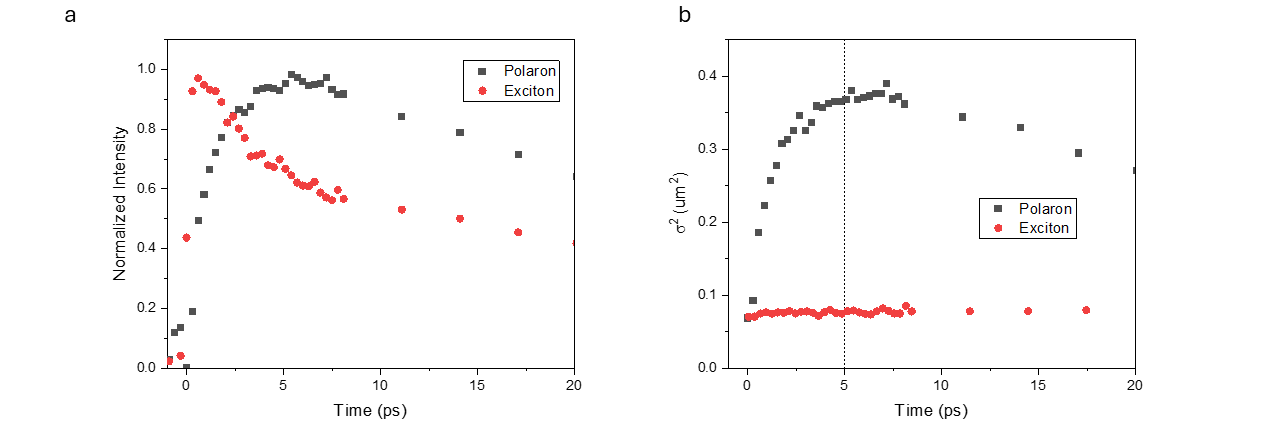}
    \caption{\textbf{Quasi-equilibrium regime of the transient optical response.}
\textbf{a,} Time dependence of the normalized differential reflectance signal for the neutral exciton at charge neutrality and the attractive exciton--Fermi polaron at a Fermi energy of 23~meV. After an initial rise associated with polaron formation, the polaron signal exhibits minimal additional decay between approximately 5~ps and 10~ps, defining a temporal window in which the population is approximately constant. Measurements within this window are therefore considered to probe a quasi-equilibrium regime.
\textbf{b,} Time dependence of the spatial second moment $\sigma^{2}$ extracted from fits to the spatial profiles of the transient response. Following the initial rise, $\sigma^{2}$ reaches a plateau that closely tracks the behavior of the integrated differential signal, consistent with a quasi-equilibrium spatial distribution during this time window.
}
    \label{Extendend Data Figure X}
\end{figure}

\begin{figure}
    \centering
    \includegraphics[width=\linewidth]{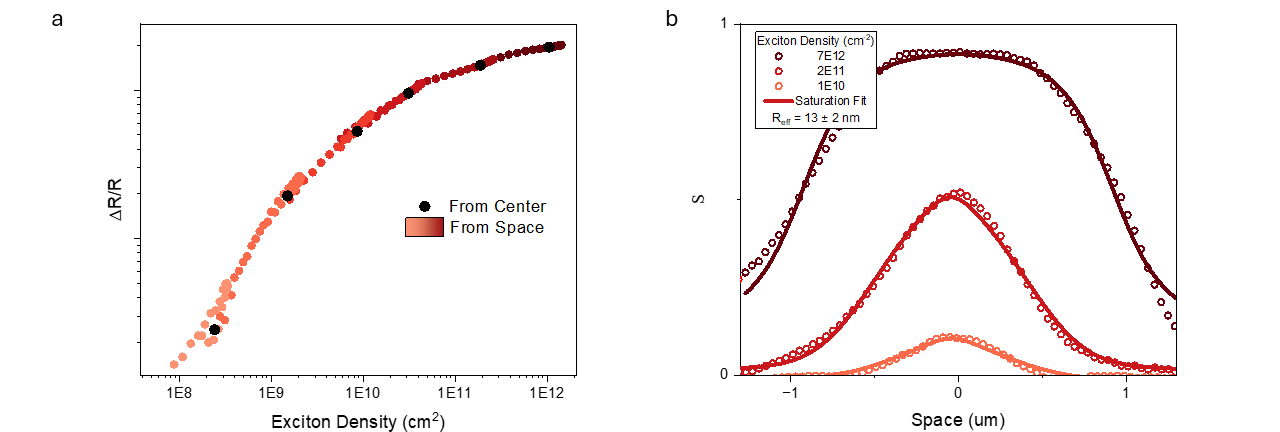}
    \caption{\textbf{Equivalence of spatial and power-dependent nonlinear response.}
\textbf{a,} Differential reflectance signal $\Delta R/R$ measured at the pump center as a function of excitation power (black) compared with the spatially resolved $\Delta R/R$ profile measured at fixed total power and normalized by the local pump intensity (red), both evaluated at a pump--probe delay of 5~ps. The close agreement between the two datasets demonstrates that the spatial dependence of the nonlinear response is fully accounted for by the local excitation density set by the pump profile, without requiring exciton or polaron transport on this timescale.
\textbf{b,} Given this equivalence, the spatial profile of the nonlinear response is modeled using a local saturation form weighted by the measured pump intensity profile, $\Delta R/R(x) = \,n_{\rm 0}(x)/\bigl[1 + \pi R_b^2 n_{\rm 0}(x)\bigr]$, where $n_{\rm 0}(x)$ is the injected excitation density above resonance proportional to the pump intensity, and $R_b$ is the effective blockade radius. The pump intensity profile $n(x)$ is obtained independently from a Voigt-function fit to the measured pump spot, such that the only free parameters in the spatial fit are $S$ and $R_b$.
\textbf{c,} The blockade radius extracted from fitting the spatial profiles is consistent, within one standard deviation, with the value obtained independently from the low-density slope of the power-dependent saturation measurements, confirming that the spatial extent of the nonlinear response reflects an interaction-induced saturation rather than transport.
}

    \label{Extendend Data Figure X}
\end{figure}

\begin{figure}
    \centering
    \includegraphics[width=\linewidth]{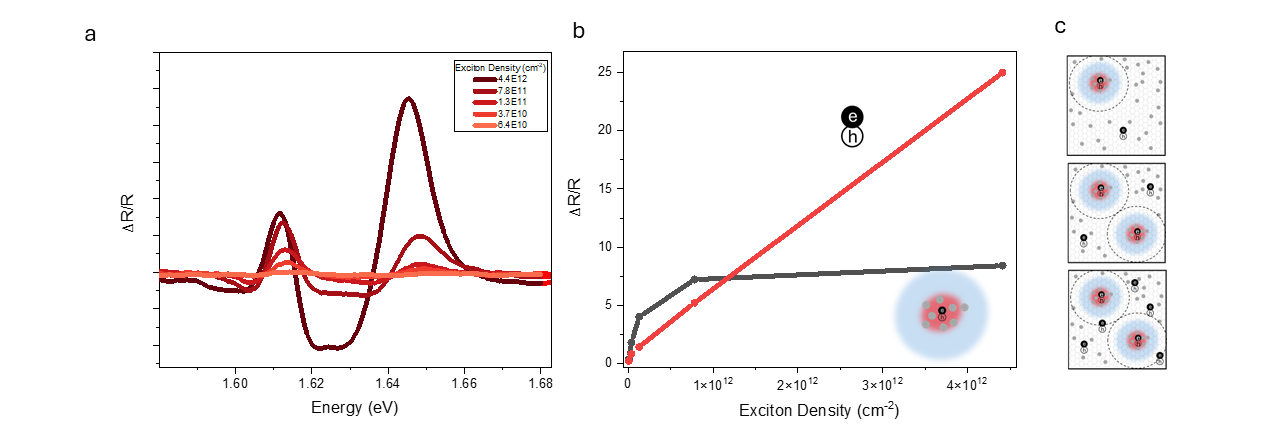}
    \caption{\textbf{Fluence-dependent transient absorption response at a Fermi energy of 5~meV.}
\textbf{a,} Fluence dependence of the transient absorption signal at a pump--probe delay of 5~ps for the exciton--Fermi polaron at a Fermi energy of 5~meV, where both attractive (AP) and repulsive (RP) polaron resonances are present in the linear-response regime.
\textbf{b,} Spectrally integrated transient absorption signal associated with the AP and RP energy ranges as a function of excitation fluence. While the RP signal increases approximately linearly with excitation density over the full range studied, the AP signal exhibits clear saturation at higher fluence. This demonstrates that the nonlinear response is not a global effect affecting all resonances equally, but instead is specific to the attractive polaron feature.
\textbf{c,} Schematic illustration of the fluence-dependent evolution of the transient absorption spectrum. At low excitation density, both AP and RP features are present and increase with fluence. With increasing excitation density, the AP resonance saturates while the RP feature continues to scale linearly, highlighting the resonance-specific nature of the nonlinear response.
}

    \label{Extendend Data Figure X}
\end{figure}

\begin{figure}
    \centering
    \includegraphics[width=\linewidth]{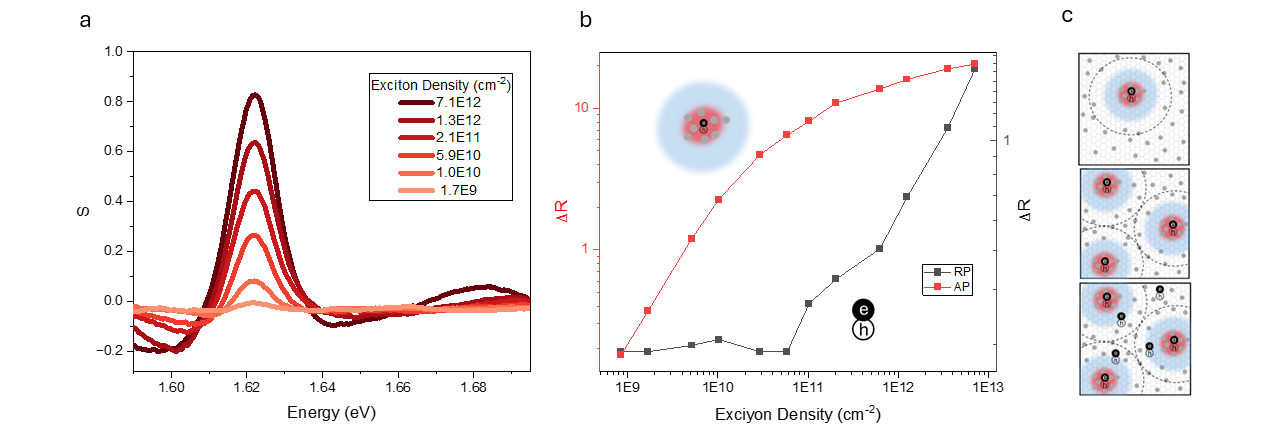}
    \caption{\textbf{Fluence-dependent transient absorption response at a Fermi energy of 23~meV.}
\textbf{a,} Fluence dependence of the transient absorption signal at a pump--probe delay of 5~ps for the attractive exciton--Fermi polaron at a Fermi energy of 23~meV, expressed in terms of the normalized saturation parameter. With increasing excitation density, the attractive polaron resonance rapidly saturates. At higher fluence, an additional differential signal emerges at higher energy, corresponding to the repulsive polaron (RP) feature.
\textbf{b,} Spectrally integrated transient absorption signal associated with the attractive polaron (AP) and repulsive polaron (RP) energy ranges as a function of excitation fluence. Following saturation of the AP resonance, spectral weight is partially redistributed toward the higher-energy RP feature, indicating preferential occupation of higher-energy polaron states at high excitation density. This spectral redistribution accounts for only a fraction of the total nonlinear response and does not explain the overall saturation behavior.
\textbf{c,} Schematic illustration of the fluence-dependent evolution of the transient absorption spectrum. At low excitation density, the response is dominated by depletion of the attractive polaron resonance. After saturation of the attractive polaron, additional population appears at higher energies associated with the repulsive polaron, reflecting a redistribution of spectral weight rather than a complete transfer of population.
}

    \label{Extendend Data Figure X}
\end{figure}

%%%%%%
\color{black}
\iffalse

\section*{Extended Data}
\fi

\end{document}